\documentclass[12pt]{article}
\usepackage{graphicx,amsmath,hyperref}
\usepackage{authblk}
\usepackage{amssymb}
\newcommand{\Rf}{{\mathbf R}}
\newcommand{\Tf}{{\mathbf T}}
\newcommand{\Pf}{{\mathbf P}}

\newcommand{\Ts}{{\mathsf T}}
\newcommand{\ps}{\mathsf{p}}

\newcommand{\g}{{\mathbf g}}

\newcommand{\1}{{\mathbf 1}}
\newcommand{\e}{{\mathbf e}}
\newcommand{\tf}{{\mathbf t}}
\newcommand{\vv}{{\mathbf v}}
\def\hybrid{\topmargin 0pt      \oddsidemargin 0pt
        \headheight 0pt \headsep 0pt
        \voffset=-0.5cm
        \textwidth 6.25in       
        \textheight 9.5in       
        \marginparwidth 0.0in
        \parskip 5pt plus 1pt   \jot = 1.5ex}
\catcode`\@=11
\def\marginnote#1{}

\newcount\hour
\newcount\minute
\newtoks\amorpm
\hour=\time\divide\hour by60
\minute=\time{\multiply\hour by60 \global\advance\minute by-\hour}
\edef\standardtime{{\ifnum\hour<12 \global\amorpm={am}%
        \else\global\amorpm={pm}\advance\hour by-12 \fi
        \ifnum\hour=0 \hour=12 \fi
        \number\hour:\ifnum\minute<10 0\fi\number\minute\the\amorpm}}
\edef\militarytime{\number\hour:\ifnum\minute<10 0\fi\number\minute}

\def\draftlabel#1{{\@bsphack\if@filesw {\let\thepage\relax
   \xdef\@gtempa{\write\@auxout{\string
      \newlabel{#1}{{\@currentlabel}{\thepage}}}}}\@gtempa
   \if@nobreak \ifvmode\nobreak\fi\fi\fi\@esphack}
        \gdef\@eqnlabel{#1}}
\def\@eqnlabel{}
\def\@vacuum{}
\def\draftmarginnote#1{\marginpar{\raggedright\scriptsize\tt#1}}
\def\draftlabel#1{{\@bsphack\if@filesw {\let\thepage\relax
   \xdef\@gtempa{\write\@auxout{\string
      \newlabel{#1}{{\@currentlabel}{\thepage}}}}}\@gtempa
   \if@nobreak \ifvmode\nobreak\fi\fi\fi\@esphack}
        \gdef\@eqnlabel{#1}}
\def\@eqnlabel{}
\def\@vacuum{}
\def\draftmarginnote#1{\marginpar{\raggedright\scriptsize\tt#1}}

\def\draft{\oddsidemargin -.5truein
        \def\@oddfoot{\sl preliminary draft \hfil
        \rm\thepage\hfil\sl\today\quad\militarytime}
        \let\@evenfoot\@oddfoot \overfullrule 3pt
        \let\label=\draftlabel
        \let\marginnote=\draftmarginnote
   \def\@eqnnum{(\theequation)\rlap{\kern\marginparsep\tt\@eqnlabel}%
\global\let\@eqnlabel\@vacuum}  }

\def\numberbysection{\@addtoreset{equation}{section}
        \def\theequation{\thesection.\arabic{equation}}}

\def\underline#1{\relax\ifmmode\@@underline#1\else
        $\@@underline{\hbox{#1}}$\relax\fi}

\def\titlepage{\@restonecolfalse\if@twocolumn\@restonecoltrue\onecolumn
     \else \newpage \fi \thispagestyle{empty}\c@page\z@
        \def\thefootnote{\fnsymbol{footnote}} }

\def\endtitlepage{\if@restonecol\twocolumn \else  \fi
        \def\thefootnote{\arabic{footnote}}
        \setcounter{footnote}{0}}  
\relax

\numberbysection
\hybrid

\newtheorem{prop}{Proposition}[section]
\newtheorem{theor}[prop]{Theorem}

\def\beq{\begin{equation}}
\def\eeq{\end{equation}}
\def\p{\partial}

\date{December 2014}

\newcommand{\la}{\lambda}
\newcommand{\al}{\alpha}
\newcommand{\be}{\beta}
\newcommand{\ga}{\gamma}

\def\res{\mathop{\hbox{Res}}\limits}

\begin{document}

\begin{titlepage}

\title{\hfill{\normalsize ITEP-TH-36/14} \\ [10mm]
{\bf Supersymmetric quantum spin chains
and classical integrable systems}}

\author[1]{Zengo Tsuboi \thanks{E-mail:  {\tt ztsuboi@yahoo.co.jp}
}}
\author[2,3,4]{Anton Zabrodin \thanks{E-mail:  {\tt zabrodin@itep.ru}
}}
\author[5,2]{Andrei Zotov \thanks{E-mail: {\tt zotov@mi.ras.ru}
}}

 \affil[1]{{\small Department of Mathematics and Statistics,
The University of Melbourne, Royal Parade, Parkville,
Victoria 3010, Australia}}
 \affil[2]{{\small ITEP, 25 B.Cheremushkinskaya, Moscow 117218, Russia}}
 \affil[3]{{\small Institute of Biochemical Physics RAS, 4 Kosygina st.,
Moscow 119334, Russia}}
 \affil[4]{{\small Laboratory of Mathematical Physics,
National Research University Higher School of Economics,
20 Myasnitskaya Ulitsa, Moscow 101000, Russia}}
 \affil[5]{{\small  Steklov Mathematical
Institute  RAS, 8 Gubkina str., Moscow 119991,  Russia}}

\maketitle

\begin{abstract}

For integrable inhomogeneous
supersymmetric spin chains (generalized
graded magnets) constructed employing
$Y(gl(N|M))$-invariant $R$-matrices
in fi\-ni\-te-\-di\-men\-si\-onal representations we introduce
the master T-operator which is a sort of generating
function for the family of commuting quantum transfer matrices.
Any eigenvalue of the master T-operator is the tau-function of the
classical mKP hierarchy. It is a polynomial in the spectral  
parameter which is identified with the 0-th time of the hierarchy.
This implies a remarkable relation between the quantum
supersymmetric spin chains and classical many-body integrable
systems of particles of the Ruijsenaars-Schneider type.
As an outcome, we obtain a system of
algebraic equations for the spectrum of the spin chain
Hamiltonians.

\end{abstract}


Journal-ref: JHEP 05 (2015) 086

\end{titlepage}

\bigskip

\newpage

\tableofcontents
\newpage

\section{Introduction}

Supersymmetric extensions of quantum integrable spin chains
were introduced in \cite{KulishandSklyanin,Kulish}.
Such models,
called graded magnets in \cite{Kulish},
are based on solutions of the graded Yang-Baxter equation
(graded analogues of quantum $R$-matrices) in the same way as
ordinary integrable spin chains are built from quantum $R$-matrices
satisfying the Yang-Baxter $RRR=RRR$ relation. An important class
of solutions are $Y(gl(N|M))$-invariant $R$-matrices taken in
finite-dimensional representations which are simple rational
functions of the spectral parameter $x$:
\beq\label{itr1}
\Rf (x)=\1 \otimes \1 +  \frac{\eta}{x}\, \mathbf{P}.
\eeq
Here $\mathbf{P}\in \mathrm{End}
\bigl (\mathbb{C}^{N|M}\otimes \mathbb{C}^{N|M}\bigr )$
is the graded permutation operator.
In this paper, we focus
on graded magnets constructed using the $R$-matrices from this class.
They are called (generalized) supersymmetric spin chains of the
XXX type, or simply susy-XXX spin chains.
Their trigonometric analogue
is known as the Perk-Schultz model \cite{Perk:1981nb}.

For our purpose we need inhomogeneous susy-XXX spin chains
or the corresponding integrable lattice models of statistical
mechanics on inhomogeneous lattices, with quasiperiodic (twisted)
boundary conditions.
Their quantum monodromy matrices $\mathbf{S}(x)$ are products of
the type
\beq\label{itr2}
\mathbf{S}(x)=
\Rf^{0L}(x-x_L)\ldots \Rf^{02}(x-x_2)\Rf^{01}(x-x_1)\,
(\g \otimes \1^{\otimes L})
\eeq
along the chain, with $x_i$ being inhomogeneity parameters
assumed to be distinct and
$\g \in GL(N|M)$ being the twist matrix
assumed to be diagonal. (The label $0$ corresponds to the
auxiliary space, where the product is taken.)
The supertrace $\Tf (x)=\mbox{str}_0 \, \mathbf{S}(x)$
of the quantum monodromy matrix in the auxiliary
space is what is called the quantum transfer matrix or the T-operator.
The graded Yang-Baxter equation implies that
the T-operators commute for all $x$, so $\Tf (x)$ is
a generating function of commuting Hamiltonians $\mathbf{H}_j$:
\beq\label{itr3}
\Tf (x)=
\mbox{str} \, \g \cdot \1 ^{\otimes L}
\,  + \sum_{j=1}^{L}\frac{\eta \,
\mathbf{H}_j}{x-x_j}\,.
\eeq
These Hamiltonians are non-local, i.e., involve
interaction between operators
on all lattice sites.
However, such models still make sense as generalized spin chains with
long-range interactions.
Alternatively,
one may prefer to keep in mind integrable lattice models of statistical
mechanics rather than spin chains as such.
In either case the final goal of the theory is diagonalization of
the T-operators.
This is usually achieved by the
Bethe ansatz method in one form or another. What we are going to do
in this paper
is to suggest an alternative approach based on a hidden connection
with classical many-body integrable systems explained below.

As an intermediate step, we need to recall
that there exists a broader family
of commuting T-ope\-ra\-tors which includes $\Tf (x)$ as a subset.
Using the fusion procedure in the auxiliary space, one can
construct an infinite family of commuting T-operators
$\mathbf{T}_{\lambda}(x)$ indexed by Young diagrams $\lambda$,
with $\mathbf{T}_{\Box}(x)=\mathbf{T}(x)$.
Following \cite{Alexandrov:2011aa},
we construct the master T-operator as their generating
(operator-valued) function of a special form. Let $\tf =
\{t_1, t_2, t_3 , \ldots \}$ be an infinite set of
auxiliary ``time variables'' and $s_\lambda (\tf )$ be the
Schur polynomials.
The master T-operator, $\Ts (x, \tf )$,
for graded magnets\footnote{A preliminary form of the master T-operator
for these models appeared in the earlier work \cite{KLT10}.}
is introduced in the same way as in \cite{Alexandrov:2011aa}
and subsequent works \cite{Alexandrov:2013aa}--\cite{Z14}:
\beq\label{itr4}
\frac{\Ts (x, \tf )}{\Ts (x, \mathbf{0})}=\sum_\lambda
\Tf _{\lambda}(x) s_\lambda (\tf ).
\eeq
By construction, this family
of operators is commutative for all
$x$, $\tf$ and can be simultaneously diagonalized:
$\Ts (x, \tf )\left |\Psi \right > =T (x, \tf )\left |\Psi \right >$.
The main fact about the master T-operator, which makes the whole
construction interesting, is that the so defined $\Ts (x, \tf )$
satisfies the bilinear identity for the classical
modified Kadomtsev-Petviashvili (mKP) hierarchy,
with $x$ being identified with the ``$0$th time'' $t_0$:
\begin{align}\label{intr5}
\oint_{{\mathcal C}} z^{(x-x')/\eta}
e^{\sum_{k\geq 1}(t_k-t'_k)z^k}
\, \Ts \bigl (x, {\mathbf t}\! -\! [z^{-1}]
\bigr )\Ts \bigl (x',{\bf t'}\! +\! [z^{-1}]\bigr )dz =0
\end{align}
for all ${\mathbf t}$, ${\mathbf t}'$, $x$, $x'$ and for
a properly chosen integration contour.
Here ${\bf t}\pm [z^{-1}]:= \bigl \{t_k \pm \frac{1}{k}\, z^{-k}\bigr \}$.
This means that any eigenvalue $T (x, \tf )$ of $\Ts (x, \tf )$ is
a tau-function of the mKP hierarchy. In this way, the
commutative algebras of susy-XXX spin
chain Hamiltonians, for all possible gradings, appear to be embedded
into the infinite integrable hierarchy of non-linear
differential-difference equations, the
mKP hierarchy \cite{Sato,DJKM83,JM83}.
This is a further development of the earlier studies
\cite{KLWZ97,ZabTMP98,KSZ08} clarifying the role
of classical integrable hierarchies in quantum integrable models.

The next step depends on
analytical properties of the eigenvalues $T(x, \tf )$ as functions
of the variable $x$.
For finite spin chains each eigenvalue
is a polynomial in $x$
of degree $L$ for any $\tf$:
\beq\label{itr6}
T (x, \tf )=e^{\mathrm{str}\, \xi (\tf , \g )}\prod_{j=1}^{L}
\bigl (x-x_j (\tf )\bigr ).
\eeq
The roots depend on the times $\tf$. At this point,
a surprising link to integrable many-body systems
of classical mechanics comes into play. Namely,
from the fact that $T(x, \tf )$ is a tau-function of the
mKP hierarchy it follows \cite{KZ95,Iliev} that
the roots $x_i$
move in the times $t_k$ as particles of the
Ruijsenaars-Schneider (RS) $L$-body system \cite{RS}
subject to the equations of motion corresponding to
the $k$th Hamiltonian ${\cal H}_k$ of the RS model.
For example, the equations of motion for the $t=t_1$ flow are
\beq\label{itr7}
\ddot x_i=-\sum_{k\neq i}
\frac{2\eta^2 \dot x_i\dot x_k}{(x_i-x_k)
\left [ (x_i-x_k)^2-\eta^2 \right ]}\,,
\qquad i=1, \ldots L.
\eeq
This link prompts to reformulate the spectral
problem for the susy-XXX spin chain Hamiltonians $\mathbf{H}_j$ in terms
of the integrable model of classical mechanics.
It is important to stress
that this quantum-classical
correspondence (the QC-correspondence)
does not depend on the choice of the
grading: all graded magnets are linked to the same RS system
of particles. The role of the QC-correspondence in
supersymmetric gauge theories and branes was discussed in
\cite{NRS,GK13,GZZ}.

The RS system is often referred to as an integrable relativistic
deformation of the famous Calogero-Moser system. Similarly to the
latter, it
admits the Lax representation, i.e. the dynamics can be translated
into isospectral deformations of a matrix
$\mathsf{Z}(\{x_i(t)\},\{\dot x_i(t)\})$ which is called the Lax
matrix.
The essence of the QC-correspondence
of integrable systems lies in the fact that the spectra of the quantum
Hamiltonians $\mathbf{H}_j$ are encoded in
the Lax matrix
$\mathsf{Z}(\{x_i(0)\},\{\dot x_i(0)\})\equiv\mathsf{Z}_0$
for the RS system at $t=0$
after the identifications $x_i(0):=x_i$ (the inhomogeneity parameters
of the spin chain) and $-\eta \, \dot x_i(0):=H_i$
(the eigenvalues of the
quantum Hamiltonians):
\beq\label{itr8}
(\mathsf{Z}_0)_{ij}=\frac{\eta \, H_i}{x_j-x_i+\eta}\,.
\eeq
Given the $x_i$'s, possible values of
the $H_i$'s are determined from the condition that the matrix
$\mathsf{Z}_0$
has a prescribed set of eigenvalues which
is a subset of $\{g_1, g_2, \ldots , g_{N+M}\}$, where
$g_i$ are elements of the
(diagonal) twist matrix $\g$, taken with certain multiplicities.
In this way the spectral problem for
the quantum Hamiltonians is reduced to a sort of inverse
spectral problem for the Lax matrix.

We give two different proofs of this remarkable correspondence.
One (indirect) is through the mKP hierarchy and its polynomial solutions.
The other proof (based on the technique
developed in \cite{GZZ}) is by a direct computation using the
description of the spectrum in terms of the (nested) Bethe ansatz
equations.
Both proofs are rather technical.
It is of value
to find a more conceptual proof.

As a corollary, computing the spectral
determinant for the RS Lax matrix, we find that the eigenvalues
of the quantum Hamiltonians for all XXX spin chains
on $L$ sites
are encoded in the following system of algebraic
equations:
\begin{align}
\sum_{1 \le i_{1} < \ldots < i_{n} \le L}
 H_{i_{1}}\dots H_{i_{n}}
\! \prod_{1 \le \alpha < \beta \le n}
\!\! \! \left(
1-\frac{\eta^{2}}{(x_{i_{\alpha}}- x_{i_{\beta}})^{2} }
\right)^{-1}
=e_{n}(g_1, \ldots , g_L), \quad n=1, \ldots , L.
 \label{intr9}
\end{align}
Here $e_n$ are elementary symmetric polynomials of
$L$ parameters $g_i$: $e_1=\sum_i g_i$, $e_2=
\sum_{i<j}g_i g_j$, etc. Identifying them with elements
of the twist matrix in a proper way
(at $L>N+M$ some $g_i$'s have to be merged), one
finds a part of the spectrum
$(H_1, \ldots , H_L)$ for a particular spin chain
among solutions to the system. Other solutions of the same system
correspond to some other spin chain. In order to find the full
spectrum of a given model, one should solve systems
of the form (\ref{intr9}), where $g_i$'s
are taken with different possible multiplicities from a given set.
Two points are worth emphasizing:
\begin{itemize}
\item
These are equations for the spectrum itself,
not for any auxiliary parameters like in the Bethe ansatz solution.
\item
This system does not depend on the grading, so the set of its
solutions contain spectra of quantum Hamiltonians for the models
with all possible gradings.
\end{itemize}
The detailed
structure of solutions and their precise correspondence with spectra
of particular spin chains is a subject of further study.

From the algebro-geometric point of view,
equations (\ref{intr9}) define a $2L$-dimensional algebraic
variety $\mathbb{S}_L$ which can be called the
{\it universal spectral variety} for spin chains of the XXX type.
It contains a comprehensive
information about spectra of spin chains
on $L$ sites based on
the $gl(N|M)$ algebras with all possible gradings.
The variety $\mathbb{S}_L$ given by equations
(\ref{intr9}) is not compact.
These equations only define its affine part embedded into the
$3L$-dimensional space with coordinates
$(H_1, \ldots , H_L; x_1, \ldots , x_L; g_1, \ldots , g_L)$.
Presumably, a proper compactification
of the universal spectral variety encodes information about the
spectra of homogeneous spin chain Hamiltonians (when $x_i\to 0$).

\paragraph{Organization of the paper.}
In Section 2 we recall the construction of the integrable
susy-XXX spin chains
starting from the quantum $R$-matrices.
For our purpose we need a fully inhomogeneous model with twisted
boundary conditions.
We introduce the non-local commuting Hamiltonians,
which are the main observables in the system, and such
attendant objects like the higher T-operators (transfer matrices).
The master T-operator is introduced in section 2.4 as their
generating function. In section 2.5 we present the bilinear
identity satisfied by the master T-operator which makes it possible
to embed the quantum stuff into the context of classical
integrable hierarchies of non-linear PDE's. In particular,
we define the classical Baker-Akhiezer function in terms of the quantum
T-operators.

In Section 3 we establish and exploit the link to the RS
$L$-body system.
The main point here is the reformulation of the eigenvalue problem
for the spin chain Hamiltonians in terms of coordinates
and velocities of the
RS particles. The Lax pair for the RS system is derived from the
poles dynamics of the Baker-Akhiezer function in Section 3.2.

Section 4 contains some details of the QC correspondence
which is based on the identification of the twist parameters with
eigenvalues of the Lax matrix for the RS model (Section 4.1).
In Section 4.3 the algebraic equations for eigenvalues of the
spin chain Hamiltonians are obtained and the notion of the
universal spectral variety is introduced.
In Section 5 we give a direct proof of the
QC correspondence, using the nested Bethe ansatz solution.
In section 6 some unsolved problems are listed.

There are also three appendices. In Appendix A we give some
technical details needed for deriving the higher T-operators
in a more or less explicit form as derivatives of supercharacters.
Appendix B is a reference source for the Hamiltonian approach to the
RS system.
Explicit examples of spin chain spectra
for small
number of sites ($L=1,2,3$) and their comparison
with solutions to the algebraic equations from
Section 4.3 are given in Appendix C.


\paragraph{Acknowledgments.}
We thank A. Alexandrov, A. Gorsky, A. Liashyk, V. Kazakov, S.
Khoroshkin, I. Kri\-che\-ver and S. Leurent for discussions. The
work of Z.\ Tsuboi is supported by the Australian Research Council.
The work of A.\ Zabrodin is supported in part by RFBR grants
14-02-00627, 12-02-91052-CNRS, 14-01-90405-Ukr and by grant
Nsh-1500.2014.2 for support of scientific schools. 
The financial support from the Government of the Russian Federation 
within the framework of the implementation of the 5-100 Programme Roadmap of 
the National Research University  Higher School of Economics is acknowledged.
The work of A.\ Zotov is supported in part by RFBR grants 14-01-00860,
15-51-52031-NNS$_a$, by the D. Zimin's fund ``Dynasty'' and by the
Program of RAS ``Basic Problems of the Nonlinear Dynamics in
Mathematical and Physical Sciences'' $\Pi$19.



\paragraph{The notation.}
Throughout the paper, we use the following notation.
\begin{itemize}
\item[]
$\mathfrak{B}$: any one of the subsets of $\{1,2,\dots, N+M\}$
with $ \mathrm{Card} (\mathfrak{B})=N$.
\item[]
$\mathfrak{F}$: the complement set
$\mathfrak{F}=\{1,2,\dots, N+M\} \setminus \mathfrak{B}$
\item[]
$\ps$: the $\mathbb{Z}_2$-grading parameter, $\ps(a)=0$ for
$a\in \mathfrak{B}$ and $\ps(a)=1$ for
$a\in \mathfrak{F}$
\item[]
$\e_{ab}$: generators
of $gl(N|M)$ identified with (super)matrix units, $(\e_{ab})_{a'b'}=\delta_{aa'}\delta_{bb'}$
\item[]
$\mathbf{v}_a$: orthonormal
basis vectors in $\mathbb{C}^{N|M}$ such that
$\e_{ab}\mathbf{v}_c=\delta_{bc}\mathbf{v}_a$,
$\ps (\mathbf{v}_a)=\ps (a)$
\item[]
$\g$: a diagonal group element of $GL(N|M)$,
$\g =\mathrm{diag}\, \bigl (g_1, \ldots , g_{N+M}\bigr )$
\item[] $\mathrm{str}$: the supertrace $\mathrm{str}\, {\cal M}=\sum_a (-1)^{\ps (a)}
{\cal M}_{aa}$,
\item[] $\mbox{sdet}$: the superdeterminant 
$\mathrm{sdet}\, {\cal M}=\exp (\mathrm{str} \log {\cal M})$,  
\item[]
$\lambda$: a Young diagram with rows $\lambda_1\geq \lambda_2 \geq
\ldots \geq \lambda_\ell >0$
\item[]
$\lambda '$: the transposed Young diagram ($\lambda$ reflected
in the main diagonal)
\item[]
$\1$: the identity element in $\mathrm{End} (\mathbb{C}^{N|M})$
or $\mathrm{End} (\mathbb{C}^{L})$
\item[]
$\mathbf{I}$: the identity operator in the tensor product spaces
like $(\mathbb{C}^{N|M})^{\otimes L}$, etc
\item[]
$\mathbf{O}$: an operator in $(\mathbb{C}^{N|M})^{\otimes L}$ or
$\mathbb{C}^{L}\otimes (\mathbb{C}^{N|M})^{\otimes L}$
\item[]
$\mathbf{O}(x)$: an operator-valued rational function of $x$
\item[]
$\mathsf{O}(x)$: an operator-valued polynomial function of $x$
\end{itemize}

The $\mathbb{Z}_2$-grading of the
$gl(N|M)$-generators $\e_{ab}$ (identified with the matrix units)
is defined as
$\ps(\e_{ab})=\ps(a)+\ps(b) \ \text{mod} \ 2$. The commutation
relations obeyed by the generators are
$\e_{ab}\e_{cd} -(-1)^{\ps(\e_{ab})\ps(\e_{cd})}\e_{cd}
\e_{ab}=\delta_{bc}\e_{ad}-(-1)^{\ps(\e_{ab})\ps(\e_{cd})}
\delta_{ad}\e_{cb}$.

Any tensor product in this paper is the
$\mathbb{Z}_2$-graded one.
Namely, for any homogeneous operators
$\{ \mathbf{A}_{i} \}_{i=1}^{4}$ the tensor product
satisfies the relation
$(\mathbf{A}_{1} \otimes \mathbf{A}_{2})
(\mathbf{A}_{3} \otimes \mathbf{A}_{4})=
(-1)^{\ps(\mathbf{A}_{2})\ps(\mathbf{A}_{3})}
(\mathbf{A}_{1}\mathbf{A}_{3}
\otimes \mathbf{A}_{2}\mathbf{A}_{4})$.

We use the notation
$\displaystyle{\overrightarrow{\prod_{j=1}^{L}}\mathbf{O}_{j}=
\mathbf{O}_{1}  \mathbf{O}_{2} \ldots \mathbf{O}_{L}}$ and
$\displaystyle{\overleftarrow{\prod_{j=1}^{L}}\mathbf{O}_{j}=
\mathbf{O}_{L} \ldots \mathbf{O}_{2} \mathbf{O}_{1}}$
for the ordered product of the
operators $\{ \mathbf{O}_{j} \}_{j=1}^{L}$.

\section{The master T-operator for supersymmetric spin chains}

\subsection{Quantum $R$-matrices}

The simplest
$Y(gl(N|M))$-invariant $R$-matrix has the form
\begin{equation}\label{R1}
\Rf(x)=\1 \otimes \1 +\frac{\eta }{x}
\sum_{a,b=1}^{K} (-1)^{\ps(b)} \e_{ab}\otimes \e_{ba}.
\end{equation}
Hereafter, we set $K\equiv N+M$. The variable $x$
is the spectral parameter. The extra parameter $\eta$ is not actually
essential because it can be eliminated by a rescaling of $x$
(unless one tends $\eta$ to $0$ as in the limit to the Gaudin model).
The $R$-matrix (\ref{R1}) is an operator in the space
${\mathbb C}^{N|M}\otimes {\mathbb C}^{N|M}$.
It can be represented as $\Rf(x)=
\1 \otimes \1 +\frac{\eta }{x}
\mathbf{P}$, where $\Pf$ is the
graded permutation operator
given by
$\Pf =\sum_{a,b=1}^{K}(-1)^{\ps(b)}
\e_{ab}\otimes \e_{ba}$.
It acts on homogeneous vectors
as follows: $\Pf \, \mathbf{x}\otimes \mathbf{y}=
(-1)^{\ps (\mathbf{x}) \ps (\mathbf{y})}
\mathbf{y}\otimes \mathbf{x}$.

Having in mind the construction
of the spin chain on $L$ sites,
one can realize the $R$-matrix as an operator in the space
${\mathbb C}^{N|M}\otimes ({\mathbb C}^{N|M})^{\otimes L}$:
\begin{equation}\label{R1}
\Rf^{0j}(x)=\1 \otimes \1^{\otimes L} +\frac{\eta }{x}
\sum_{a,b=1}^{K} (-1)^{\ps(b)}
\e_{ab}\otimes \e_{ba}^{(j)},
\end{equation}
where $\e_{ba}^{(j)}:=\1^{\otimes (j-1)}
\otimes \e_{ba} \otimes \1^{\otimes (L-j)}$
for $j \in \{1,2,\dots, L \}$.\footnote{The definition
of the graded tensor
product implies that
$\e_{ab}^{(i)}\e_{cd}^{(j)}=
(-1)^{(\ps (a)+\ps (b))(\ps (c)+\ps (d))}
\e_{cd}^{(j)}\e_{ab}^{(i)}$ for $i\neq j$.}
The first space ${\mathbb C}^{N|M}$ labelled by
the index $0$ is called the auxiliary space
while the space ${\mathcal V}=({\mathbb C}^{N|M})^{\otimes L}$ is the
quantum space of the model.
The matrix elements $(\Rf^{0j}(x))_{ab}$ of the operator
$\Rf^{0j}(x)$ with respect to the auxiliary space
are operators in the quantum space.
They are defined by
$$
\Rf^{0j}(x) (\vv_{a} \otimes \1^{\otimes L}) =
\sum_{b=1}^{K} (\vv_{b} \otimes \1^{\otimes L})
( 1 \otimes (\Rf^{0j}(x))_{ba} )
=\sum_{b=1}^{K} \vv_{b}
 \otimes (\Rf^{0j}(x))_{ba},
$$
where $\vv_{a}$ are orthonormal basis vectors in ${\mathbb C}^{N|M}$.
From (\ref{R1}) we obtain:
\begin{align}
 (\Rf^{0j}(x))_{ab} = \delta_{ab}
\1^{\otimes L}  +
(-1)^{\ps(a) \ps(b)} \frac{\eta }{x}
\, \e_{ba}^{(j)},
\end{align}
where we have used
$$(\e_{ac}\otimes \e_{ca}^{(j)})
(\vv_{b} \otimes \1^{\otimes L})
=(-1)^{\ps(\e_{ca}^{(j)})\ps(\vv_{b})}\e_{ac} \vv_{b} \otimes \e_{ca}^{(j)}
=(-1)^{(\ps(c)+\ps(a))\ps(b)}
\delta_{cb} \vv_{a} \otimes \e_{ca}^{(j)}.$$
For example, in the $gl(1|1)$-case
with the grading $(\ps(1),\ps(2))=(0,1)$
the block matrix representation reads
\begin{align}
\Rf^{0j}(x) =
\begin{pmatrix}
\mathbf{I} + \frac{\eta }{x} \e_{11}^{(j)} &
\frac{\eta }{x} \, \e_{21}^{(j)}\\  &\\
\frac{\eta }{x} \, \e_{12}^{(j)} & \mathbf{I} -
\frac{\eta }{x} \e_{22}^{(j)}
\end{pmatrix}.
\end{align}
Here $\mathbf{I}\equiv \1^{\otimes L}$.
Below we will keep the notation $\1$ for identity elements of
$\mathrm{End}(\mathbb{C}^{N|M})$ and $\mathrm{End}(\mathbb{C}^{L})$
and will
often write $\mathbf{I}$ for the identity
operator in any other spaces involved.

One may also extend the definition of the graded permutation
to any two tensor factors of the space ${\cal V}$:
\beq\label{grperm}
\Pf_{ij} =\sum_{a,b=1}^{K}(-1)^{\ps(b)}
\e_{ab}^{(i)} \e_{ba}^{(j)}.
\eeq
On tensor products of the basis vectors it acts as follows (here $i <j$):
\begin{multline} \label{grperm1}
\Pf_{ij}
(\vv_{a_{1}} \otimes \dots \otimes \vv_{a_{i}} \otimes
\dots \otimes \vv_{a_{j}} \otimes  \dots
\otimes \vv_{a_{L}})
\\
=
(-1)^{\ps(a_{i}) +
(\ps(a_{i})+\ps(a_{j})) \sum_{k=i}^{j-1}\ps(a_{k}) }
(\vv_{a_{1}} \otimes \dots \otimes \vv_{a_{j}} \otimes
\dots \otimes \vv_{a_{i}} \otimes  \dots
\otimes \vv_{a_{L}}).
\end{multline}

The aforesaid is related to the $R$-matrix in the
evaluation representation of $Y(gl(N|M))$ based on
the vector
representation of $gl(N|M)$.
More generally, one can consider other
irreducible finite-dimensional representations.
In this paper we shall restrict our consideration to
the class of covariant tensor representations
\cite{BB81,BMR83,GZ90}.
Any partition $\lambda$ (identified with the Young diagram)
labels a covariant representation $\pi_{\lambda}$ of the
universal enveloping algebra $U(gl(N|M))$ if
$\lambda_{N+1}\leq M$.
For any such representation
one can construct an $R$-matrix
acting in the tensor product of two spaces, one of which being
the representation space $V_{\lambda}$ where the representation
$\pi_{\lambda}$ is realized, while the other one is still
$\mathbb{C}^{N|M}$.
We distinguish two $R$-matrices of this type
depending on the order of the spaces.
One is the $R$-matrix with the auxiliary space
$V_{\lambda}$.
It has the form
\begin{align}\label{Rlam}
\Rf_{\lambda}(x)= \1 \otimes \1 +
 \frac{\eta }{x}\sum_{a,b=1}^{K}  (-1)^{\ps(b)}
  \pi_{\lambda}(\e_{ab}) \otimes \e_{ba}.
\end{align}
The auxiliary space of the other one is
$\mathbb{C}^{N|M}$:
\begin{align}\label{Rlam-1}
\Rf^{\lambda}(x)= \1 \otimes \1 +
 \frac{\eta }{x}\sum_{a,b=1}^{K}  (-1)^{\ps(b)}
  \e_{ab} \otimes \pi_{\lambda}(\e_{ba}).
\end{align}
Clearly, $\Rf_{\Box}(x)=\Rf^{\Box}(x)=\Rf (x)$.

It is convenient to denote
\beq\label{Plam}
{\bf P}_{\lambda}^{0j}=\sum_{a,b=1}^{K}  (-1)^{\ps(b)}
  \pi_{\lambda}(\e_{ab}) \otimes \e_{ba}^{(j)}, \qquad
  {\bf P}^{\lambda}_{0j}=\sum_{a,b=1}^{K}  (-1)^{\ps(b)}
  \e_{ab} \otimes \pi_{\lambda}(\e_{ba}^{(j)}),
\eeq
then the $R$-matrix acting non-trivially in the tensor product of the
auxiliary space $V_{\lambda}$ and the $j$th space $\mathbb{C}^{N|M}$ is
$\displaystyle{\Rf_{\lambda}^{0j}(x)= {\bf I}+
\frac{\eta }{x}\, {\bf P}_{\lambda}^{0j}}$ while
the $R$-matrix acting non-trivially in the tensor product of the
auxiliary space $\mathbb{C}^{N|M}$ and the $j$th space
$V_{\Lambda ^{(j)}}$ is
$\displaystyle{\Rf^{\Lambda ^{(j)}}_{0j}(x)= {\bf I}+
\frac{\eta }{x}\, {\bf P}^{\Lambda ^{(j)}}_{0j}}$.
The $R$-matrix $\Rf^\lambda (x)$ obeys the graded Yang-Baxter
equation
\beq\label{YB1}
\Rf ^{\Box}_{12}(x_1 \! -\! x_2)
\Rf ^{\lambda}_{13}(x_1 \! -\! x_3)
\Rf ^{\lambda}_{23}(x_2 \! -\! x_3)=
\Rf ^{\lambda}_{23}(x_2 \! -\! x_3)
\Rf ^{\lambda}_{13}(x_1 \! -\! x_3)
\Rf ^{\Box}_{12}(x_1 \! -\! x_2)
\eeq
and possesses the invariance property
\beq\label{YB2}
\pi_{\Box}(\g) \otimes \pi_{\lambda}(\g) \, \Rf ^{\lambda}(x)=
\Rf ^{\lambda}(x)\, \pi_{\Box}(\g) \otimes \pi_{\lambda}(\g)
\eeq
valid for any $\g$.
The graded Yang-Baxter equation for the $R$-matrix
$\Rf_\lambda (x)$ reads
\beq\label{YB3}
\Rf _{\lambda , \mu}^{12}(x_1 \! -\! x_2)
\Rf _{\lambda}^{13}(x_1 \! -\! x_3)
\Rf _{\mu}^{23}(x_2 \! -\! x_3)=
\Rf _{\mu}^{23}(x_2 \! -\! x_3)
\Rf _{\lambda}^{13}(x_1 \! -\! x_3)
\Rf _{\lambda , \mu}^{12}(x_1 \! -\! x_2),
\eeq
where $\Rf _{\lambda , \mu}^{12}(x)\in \mathrm{End}
\bigl (V_{\lambda}\otimes V_{\mu}\bigr )$ is a yet more general
$R$-matrix. Its explicit form is complicated.
The invariance property for the $\Rf_\lambda (x)$ is similar
to (\ref{YB2}) with the opposite order of the tensor factors.

\subsection{Inhomogeneous susy-XXX chains}

Here we construct, using the $R$-matrices from the previous
subsection, the
inhomogeneous integrable susy-XXX spin chains with twisted
boundary conditions.

\subsubsection{T-operators, non-local Hamiltonians and integrals
of motion}

Let $\g \in GL(N|M)$ be a group element represented
by a diagonal (super)matrix $\g =\mbox{diag}\, (g_1, g_2, \ldots , g_{K})=
\displaystyle{\sum_{a=1}^{K}g_a \e_{aa}}$.
We call it the twist matrix with the twist parameters $g_i$.
It is used for the construction of an integrable spin chain
with twisted boundary conditions.
The T-operator (the transfer matrix) of
the inhomogeneous spin chain with twisted boundary conditions
is defined by
\begin{align}
\Tf (x)=\mathrm{str}_{0}
\Bigl (\Rf ^{0L}(x-x_{L})
\ldots \Rf ^{02}(x-x_{2}) \Rf ^{01}(x-x_{1})
\, (\g \otimes \mathbf{I})\Bigr ),
 \label{Top0}
\end{align}
where $x_1, x_2, \ldots , x_L$ are inhomogeneity parameters.
We assume that they are in general position meaning that
$x_i\neq x_j$ and $x_i\neq x_j \pm \eta$ for all
$i\neq j$. As is known, the Yang-Baxter equation implies that
the T-operators with fixed inhomogeneous and twist
parameters commute:
$[\Tf (x), \, \Tf (x')]=0$
for any $x, x'$.

The dynamical variables of the model
(which we call ``spins'' in analogy with the rank 1 case)
are vectors in the vector representation of
$gl(N|M)$ realized in the spaces $\mathbb C^{N|M}$ attached to each site.
One can define a set of non-local commuting Hamiltonians $\mathbf{H}_{j}$
as residues of $\Tf (x)$ at $x=x_{j}$:
\begin{align}
\Tf (x)={\bf I} \, \mathrm{str} \, \g \,  +
\sum_{j=1}^{L}
\frac{\eta \mathbf{H}_{j}}{x-x_{j}}.
\label{T1ex}
\end{align}
In general, the Hamiltonians $\mathbf{H}_{j}$ imply
a long-range interaction involving all spins in the
chain (cf. \cite{HKW92}).
Their explicit form is
\begin{align}\label{hamham}
\mathbf{H}_{j}&=
\overleftarrow{\prod_{k=1}^{j-1}}
\left(\mathbf{I}+\frac{\eta \Pf_{kj}}{x_{j}-x_{k}} \right)
\g^{(j)}
\overleftarrow{\prod_{k=j+1}^{L}}
\left(\mathbf{I}+\frac{\eta \Pf_{jk}}{x_{j}-x_{k}} \right) \\
&=
\sum_{I \subseteq \{1,2,\dots, L\} \setminus \{j\}}
\eta^{|I|}
\left (\prod_{k \in I}\frac{1}{x_{j}-x_{k}}\right )
\left (\overleftarrow{\prod _{k \in I,  k<j} }
\Pf_{kj} \right )\,
\g^{(j)}\,
\left (\overleftarrow{\prod_{k\in I, k>j}} \Pf_{jk} \right ),
\end{align}
where $\g^{(j)}:=\1^{\otimes (j-1)}
\otimes \g \otimes \1^{\otimes (L-j)}$
and $\Pf_{ij}$ is the graded permutation operator (\ref{grperm}).
In the second line, the sum is taken over all
subsets $I$ of the set $\{1,2,\dots, L\} \setminus \{j\}$
including the empty one; $|I|\equiv \mathrm{Card}\, I$.

In addition to the Hamiltonians $\mathbf{H}_{j}$, there are
other integrals of motion.
It is easy to see that the operators
\begin{align}\label{Ma}
{\mathbf M}_a =\sum_{j=1}^L \e_{aa}^{(j)},
\end{align}
referred to as {\it weight operators},
commute with the $\mathbf{H}_i$'s: $[\mathbf{H}_{j}, {\mathbf M}_a]=0$.
Therefore, the eigenstates
of the Hamiltonians can be classified according to the
eigenvalues $(M_1, \ldots , M_K)$ of the weight operators referred to as
{\it weights}. For example,
in the $gl(2)$-case, $M_1$ and $M_2$ are the numbers of spins with
positive and negative $z$-projections respectively.

Let
$\displaystyle{
\mathcal{V}=(\mathbb{C}^{N|M})^{\otimes L}=\!\!\!\!
\bigoplus_{\phantom{aaa}M_1, \ldots , M_{K}}
\!\! \mathcal{V}(\{M_a\})}
$
be the ``weight decomposition'' of the quantum space into the direct sum
of {\it weight spaces} which are eigenspaces of the weight operators with
the eigenvalues $M_a \in \mathbb{Z}_{\geq 0}$, $a=1, \ldots , K$.
Then any eigenstate of the $\mathbf{H}_{j}$'s belongs to some weight space
$\mathcal{V}(\{M_a\})$.
The dimension of the weight space $\mathcal{V}(\{M_a\})$ is given by
$$
\mbox{dim}\, \mathcal{V}(\{M_a\})=\frac{L!}{M_1! \ldots M_K!}\,.
$$
In particular, let $1\leq a_0 \leq K$
be some fixed index, then the space with
$M_a=L \delta_{aa_0}$ is one-dimensional. It is spanned by the vector
$\vv_{a_0} \otimes \ldots \otimes \vv_{a_0}$ which is an eigenvector
of the Hamiltonians (\ref{hamham}). Indeed, using (\ref{grperm1})
in the particular case
$a_{1}=\ldots =a_{L}=a_0$, one can see that
\begin{align}
\mathbf{H}_{j} (\vv_{a_0} \otimes \dots \otimes \vv_{a_0})
=g_{a_0} \prod_{k=1, \ne j}^{L}
\left(1+\frac{(-1)^{\ps(a_0) }\eta}{x_{j}-x_{k}} \right)
(\vv_{a_0} \otimes \ldots \otimes \vv_{a_0}).
\end{align}

The weight operators are not all independent.
Since $\displaystyle{\sum_a \e_{aa}=\1}$, we have
$\displaystyle{\sum_a {\mathbf M}_a=L\, \mathbf{I}}$ and hence
$\displaystyle{\sum_a M_a =L}$. Note also that
\begin{align}\label{HandM}
\sum_{j=1}^L \mathbf{H}_j =
\sum_{j=1}^L \mathbf{g}^{(j)}=
\sum_{a=1}^{K} g_a  \mathbf{M}_a,
\end{align}
so the model has $L\! +\! N\! +\! M\! -\! 1$ independent
commuting integrals of motion.

For completeness, we give here the definition of the
T-operator for a more general inhomogeneous spin chain model
with the quantum space $\bigotimes_{j=1}^L V_{\Lambda^{(j)}}$
and the auxiliary space $\mathbb{C}^{N|M}$.
The spin chain is defined by the following data:
 \begin{itemize}
 \item The number of sites, $L$, and the inhomogeneity parameters $x_i$ at
each site;
 \item Covariant tensor
representations of $gl(N|M)$ indexed by the Young diagrams
\beq\label{s02}
\Lambda^{(j)}=(\Lambda^{(j)}_1, \ldots ,\Lambda^{(j)}_K)
\in (\mathbb{Z}_{\geq 0})^{K}\,,\qquad
\Lambda^{(j)}_1\geq  \Lambda^{(j)}_2 \geq \ldots \geq \Lambda^{(j)}_K
\geq 0
\eeq
 assigned to each site $j=1, \ldots ,L$;
  \item Elements of the
  diagonal twist matrix $\g =\hbox{diag}(g_1,...,g_K)$ (the twist parameters).
   \end{itemize}
   The T-operator
\begin{align}
\Tf ^\Lambda (x)=\mathrm{str}_{0}
\Bigl (\Rf ^{\Lambda^{(L)}}_{0L}(x-x_{L})
\ldots \Rf ^{\Lambda^{(2)}}_{02}(x-x_{2}) \Rf ^{\Lambda^{(1)}}_{01}(x-x_{1})
\, (\g \otimes \mathbf{I})\Bigr ).
 \label{Top01}
\end{align}
acts in the space $\bigotimes_{j=1}^L V_{\Lambda^{(j)}}$.
One may also introduce a set of
Hamiltonians $\mathbf{H}^{\Lambda}_{j}$ in the way
similar to (\ref{T1ex}):
\begin{align}
\Tf ^{\Lambda} (x)={\bf I} \, \mathrm{str} \, \g \,  +
\sum_{j=1}^{L}
\frac{\eta \mathbf{H}^{\Lambda}_{j}}{x-x_{j}}.
\label{T1ex-1}
\end{align}
Our main objects of interest will be the T-operator $\Tf (x)$
and the Hamiltonians $\mathbf{H}_j$
of the model with vector representations at the sites
corresponding to the choice $\Lambda ^{(j)}=(1, 0, \ldots , 0)$
for all $j=1,\ldots , L$.

\subsubsection{Diagonalization of the T-operator via
Bethe ansatz}

The algebraic form of
the nested Bethe ansatz technique for the twisted $Y(gl(n))$ inhomogeneous
spin chain \cite{KR} can be naturally extended to the $Y(gl(N|M))$-case
\cite{Kulish,KSZ08,BR,Saleur99,RS07}. The T-operators and the Hamiltonians
$\mathbf{H}_j^{\Lambda}$ can be diagonalized by this
method. Although in what follows
we need only the result
for the choice $\Lambda ^{(j)}=(1, 0, \ldots , 0)$
\footnote{The Bethe ansatz
for trigonometric models closely related to this case
was discussed in \cite{Sc83}.}, we
give here the general result for future references.

Eigenvalues of the
T-operator $\mathbf{T}^{\Lambda}(x)$ are given by
 \beq\label{s06a}
  \begin{array}{c}
 \displaystyle{
\mathrm{T}^{\Lambda}(x)= \sum\limits_{b=1}^K (-1)^{\ps(b)}g_b }
\\ \ \\
\displaystyle{ \times\prod\limits_{k=1}^L
\frac{x\!-\!x_k\!+\!(-1)^{\ps(b)}\eta\, \Lambda^{(k)}_b}{x-x_k}
\prod\limits_{\ga=1}^{L_{b\!-\!1}}\frac{x\!-\!\mu_\ga^{b\!-\!1}
\!+\!(-1)^{\ps(b)}\eta}{x-\mu_\ga^{b\!-\!1}}
\prod\limits_{\ga=1}^{\,L_{b}}\frac{x\!-\!\mu_\ga^{b}
\!-\!(-1)^{\ps(b)}\eta}{x-\mu_\ga^{b}}\,,
 }
 \end{array}
  \eeq
The corresponding eigenvalues of the
Hamiltonians (\ref{T1ex}) are as follows:
 \beq\label{s061a}
  \begin{array}{c}
 \displaystyle{
H_{\Lambda, \, i} ={\eta}^{-1}\res\limits_{x=x_i} \mathrm{T}^{\Lambda}(z)=
\sum\limits_{b=1}^K \Lambda^{(k)}_b g_b
 }
\\ \ \\
\displaystyle{ \times \prod\limits_{k\neq i}^L
\frac{x_i\!-\!x_k\!+\!(-1)^{\ps(b)}
\Lambda^{(k)}_b \eta}{x_i-x_k}
\prod\limits_{\ga=1}^{L_{b\!-\!1}}\frac{x_i\!-\!\mu_\ga^{b\!-\!1}
\!+\!(-1)^{\ps(b)}\eta}{x_i-\mu_\ga^{b\!-\!1}}
\prod\limits_{\ga=1}^{\,L_{b}}\frac{x_i\!-\!\mu_\ga^{b}
\!-\!(-1)^{\ps(b)}\eta}{x_i-\mu_\ga^{b}}\,.
 }
 \end{array}
  \eeq
It is convenient to set $L_0=L_K=0$.
The parameters $\mu^b_\al$ with
\beq\label{s04}
 \begin{array}{c}
 \displaystyle{
 \al=1,...,L_b\,,\ \ b=1,...,K-1\,,\ \ L\geq L_1\geq
 L_2\geq ... \geq L_{K-1}\geq 0
 }
 \end{array}
  \eeq
are Bethe roots. They satisfy
   the system of Bethe equations which are equivalent to the conditions
  \beq\label{s07a}
 \begin{array}{c}
 \displaystyle{
 \res\limits_{x=\mu_\al^b}\mathrm{T}^{\Lambda}(x)=0 \qquad
 \mbox{for all} \;\; \al=1,...,L_b\,,\ \ b=1, \ldots ,K-1\,.
 }
 \end{array}
  \eeq
The Bethe equations have the form:
 \beq\label{s08a}
  \begin{array}{c}
  \displaystyle{
{g_b}\,\prod\limits_{k=1}^L \frac{\mu^b_\be -x_k +
(-1)^{\ps (b)}\, \Lambda^{(k)}_b \eta}{\,\,
\mu^b_\be-x_k+(-1)^{\ps(b+1)}\! \Lambda^{(k)}_{b+1}\eta}
 \,\prod\limits_{\ga=1}^{L_{b-1}}\frac{\mu^b_\be-\mu_\ga^{b-1}+
 (-1)^{\ps(b)}\eta}{\mu^b_\be-\mu_\ga^{b-1} }
 }
 \\ \ \\
  \displaystyle{=\,\,
 {g_{b+1}}\,
\prod\limits_{\ga\neq \be}^{L_{b}}
\frac{\mu^b_\be-\mu_\ga^{b}+(-1)^{\ps(b+1)}\eta}{\mu^b_\be-\mu_\ga^{b}-(-1)^{\ps(b)}\eta}
\,\prod\limits_{\ga=1}^{L_{b+1}}\frac{\mu^b_\be-\mu_\ga^{b+1}-
(-1)^{\ps(b+1)}\eta }{\mu^b_\be-\mu_\ga^{b+1}}}\,.
\end{array}
  \eeq
Later we will specify
these general formulae for the highest weights
 \beq\label{s09}
 \begin{array}{c}
 \displaystyle{
\Lambda^{(j)}=(1,0,...,0)\ \quad \text{for all} \quad
 j=1,\ldots ,L,\ \ \hbox{i.e.,}\ \
\Lambda^{(j)}_b=\delta_{b1}\,.
 }
 \end{array}
  \eeq
With this choice the first product in the l.h.s. of (\ref{s08a})
disappears for $b\geq 2$.

\subsection{The higher T-operators}

The $R$-matrix (\ref{Rlam})
allows one to construct a family
of T-operators with the more general auxiliary space:
\begin{align}
\Tf_{\lambda }(x)=\mathrm{str}_{V_\lambda}
\Bigl (\Rf_{\lambda}^{0L}(x-x_{L})
\ldots \Rf_{\lambda}^{02}(x-x_{2}) \Rf_{\lambda}^{01}(x-x_{1})
\, (\pi_{\lambda}(\g) \otimes \mathbf{I})\Bigr ).
 \label{Top}
\end{align}
Obviously, $\Tf _{\Box}(x)$ coincides with
the T-operator $\Tf (x)$ introduced previously.
The T-operators with fixed inhomogeneous and twist
parameters commute with $\Tf (x)$ and
among themselves:
\beq\label{comm}
[\Tf_{\lambda }(x), \, \Tf_{\mu }(x')]=0
\eeq
for any $x, x', \lambda , \mu$. At $\lambda =\emptyset$
we put $\Tf_{\emptyset}(x)$ equal to
the identity operator: $\Tf_{\emptyset}(x)={\bf I}$.

It is clear from (\ref{Rlam}) and (\ref{Top})
that at $L=0$ (the empty quantum space) as well as in the limit
$x\to \infty$ for any $L$ the
T-operators become equal to the supercharacters
$\chi_{\lambda}(\g )=\mbox{str}_{V_{\lambda}}\g$.
In what follows we also need the next-to-leading term
of the expansion of $\Tf_{\lambda }(x)$
as $x\to \infty$. From the definition (\ref{Top})
one obtains the expansion
\beq\label{ntl}
\Tf_{\lambda}(x)=\chi_{\lambda}(\g ) \, {\bf I}+
\frac{\eta}{x}\sum_{j=1}^{L}\sum_{a,b}(-1)^{\ps (b)}
\frac{\p \chi_{\lambda}(
e^{\varepsilon \e _{ab}}\g )}{\p \, \varepsilon}
\Bigr |_{\varepsilon =0}\!
\e_{ba}^{(j)}+
O(1/x^2).
\eeq
Indeed, we have
$\displaystyle{
\Tf_{\lambda}(x)=\chi_{\lambda}(\g ) \, {\bf I}+
\frac{\eta}{x}\sum_{j=1}^{L}
\mbox{str}_{V_\lambda}\left (
{\bf P}_{\lambda}^{0j}\pi_{\lambda}(\g )\right )
+ O(1/x^2)}$ which is converted to the form (\ref{ntl})
by the following chain of equalities:
$$\displaystyle{
\mbox{str}_{V_\lambda}\left (
{\bf P}_{\lambda}^{0j}\pi_{\lambda}(\g )\right )=\!
\sum_{a,b}(-1)^{\ps (b)}
\mbox{str}_{V_\lambda} \pi_{\lambda}(\e_{ab}\g )\,
\e_{ba}^{(j)}=\!
\sum_{a,b}(-1)^{\ps (b)}\frac{\p}{\p \varepsilon}\Bigl [
\mbox{str}_{V_\lambda} \pi_{\lambda} \! \left (e^{\varepsilon \e_{ab}}
\g \right )\Bigr ]\Bigr |_{\varepsilon =0}\! \e_{ba}^{(j)}}.
$$
In fact the following explicit expression for the
T-operator in terms of the supercharacters is available:
\beq\label{ntl3}
\begin{array}{lll}
\Tf_{\lambda}(x)&=&\displaystyle{\sum_{l=0}^{L}\eta^l
\!\!  \sum_{i_1 <  \ldots  <  i_l}^L
\!\!\! \sum_{{\tiny \mbox{
$\begin{array}{c}\tiny{a_1, \ldots , a_l}\\
b_1, \ldots , b_l
\end{array}$
}}}\!\!
\overrightarrow{\prod_{\alpha =1}^{l}}\left (
\frac{(-1)^{\ps (b_{\alpha})}
\e _{b_{\alpha}a_{\alpha}}^{(i_\alpha )}}{x-x_{i_\alpha}}
\, \frac{\partial}{\partial \varepsilon_{\alpha}} \right )\!\!
\,
\chi_{\lambda}\bigl (e^{\varepsilon_l \e_{a_lb_l}}\! \ldots
e^{\varepsilon_1 \e_{a_1b_1}} \g
\bigr )
\Biggm |_{\varepsilon_\alpha =0}}
\\ && \\
&=&
\displaystyle{
\overrightarrow{\prod_{l =1}^{L}}\left (\mathbf{I}+
\eta \sum_{a_l, b_l}\frac{(-1)^{\ps (b_l)}
\e^{(l)}_{b_l a_l}}{x-x_l}\, \frac{\partial}{\partial \varepsilon_{l}}
\right )
\,
\chi_{\lambda}\bigl (e^{\varepsilon_l \e_{a_lb_l}}\! \ldots
e^{\varepsilon_1 \e_{a_1b_1}} \g
\bigr )
\Biggm |_{\varepsilon_l =0}}.
\end{array}
\eeq
The summation over each $a_{\alpha}$ and $b_{\alpha}$ runs from
$1$ to $K=N+M$.
The derivation of (\ref{ntl3}) is sketched in Appendix A
\footnote{
In parentheses in the second line one can recognize
the (graded) co-derivative operator \cite{KV07} which is a version
of the matrix derivative. It
proved to be a valuable technical tool for the proof of the CBR
identities and the master T-operator construction.
However, in this paper we do not use the co-derivative
explicitly.}.

As is known \cite{BB81}, supercharacters of the covariant
tensor representations are expressed by the same formulas
as for ordinary groups except for traces replaced by supertraces.
In particular, the supercharacters are known
to satisfy the Jacobi-Trudi identities \cite{Macdonald}
which have
superficially
the same form as the ones
for the usual characters:
\begin{equation}\label{JT}
\chi_{\lambda }(\g )
=\det_{1 \le i,j \le \lambda^{\prime}_{1}}
\chi_{\lambda_{i}-i+j}(\g )=\det_{1 \le i,j \le \lambda_{1}}
\chi^{\lambda^{\prime}_{i}- i+j} (\g ).
\end{equation}
Here $\chi_k:=\chi_{(k)}$ (respectively,
$\chi^k :=\chi_{(1^k)}$) is the character corresponding to the
one-row (respectively, one-column) diagram of length $k$.

There exist analogues of these identities for the
T-operators depending on the spectral parameter. These are
the Cherednik-Bazhanov-Reshetikhin (CBR) determinant formulas
sometimes called the quantum Jacobi-Trudi identities:
\begin{align}\label{CBR0}
\Tf_{\lambda }(x)=\det_{1 \le i,j \le \lambda^{\prime}_{1}}
\Tf_{\lambda_{i}-i+j}\bigl (x\! -\! (j\! -\! 1)\eta \bigr )=
\det_{1 \le i,j \le \lambda_{1}}
\Tf^{\lambda^{\prime}_{i}- i+j}\bigl (x\! +\! (j\! -\! 1)\eta \bigr ).
\end{align}
The determinants are well-defined
because all the T-operators commute.
Similarly to (\ref{JT}),
$\Tf_{k}(x):=\Tf_{(k)}(x)$ and $\Tf^{k}(x):=\Tf_{(1^k)}(x)$
are the T-operators corresponding to the one-row and
one-column diagrams respectively.
There are the following
``boundary conditions'' for them:
$\Tf_{k}(x)=\Tf^{k}(x)=0$ if $k<0$ for $N,M\neq 0$;
$\Tf_{k}(x)=0 $ if $k<0$ or $k>M$ at $N=0$;
$\Tf^{k}(x)=0 $ if $k<0$, or $k>N$ at $M=0$.

For models based on $Y(gl(N))$-invariant $R$-matrices,
the quantum Jacobi-Trudi identities follow from
resolutions of modules for the Yangian $Y(gl(N))$ \cite{Cherednik89}.
In the physical literature, they appeared in \cite{BR90} for
$gl(N)$ (see also \cite{KNS}),
 in \cite{Tsuboi:1997iq} for $gl(N|M)$
and in \cite{Hegedus09,GKT10} for some infinite dimensional representations
 in the context of AdS/CFT correspondence.

Remarkably, their general form does not depend on the grading,
although their matrix elements and
eigenvalues of the T-operators do.
In fact the assertion that
the T-operators for supersymmetric integrable models satisfy
the quantum Jacobi-Trudi identities
was a conjecture in \cite{Tsuboi:1997iq}.
A direct proof of this fact
was given in \cite{KV07} for models based on the
$Y(gl(N|M))$-invariant
$R$-matrices, where each ``spin'' of the chain
was assumed to be in the vector
representation\footnote{
There was a minor gap in the proof given in \cite{KV07} which
was filled in the appendix of \cite{Alexandrov:2011aa}
 for the $gl(N)$ case. The proof for
 $gl(N|M)$ is similar.}.
A proof which is independent of the quantum space
is only available \cite{BT08}
for models based on the $q$-deformed algebra
$U_{q}(sl(2|1))$ in the case of rectangular Young diagrams.

The eigenvalues of the T-operators (\ref{Top}) are
rational functions of $x$ with $L$ poles. Another
normalization, where they are polynomials in $x$ of degree
$L$, is also convenient and even preferable
for the link to classical integrable hierarchies.
The polynomial form of the
T-operators is obtained as
follows:
\begin{align}
\Ts_{\lambda}(x)=
\prod_{j=1}^{L}(x-x_{j})\,
\Tf_{\lambda}(x).
 \label{Top-poly}
\end{align}
In particular for $\lambda =\emptyset $, we have
$\displaystyle{\Ts_{\emptyset }(x)=\prod_{j=1}^{L}(x-x_{j})}$.
The CBR-formulas (\ref{CBR0}) in the polynomial normalization
acquire the form
\begin{align}\label{CBR}
\Ts_{\lambda }(x)=\Bigl (\prod_{k=1}^{\lambda^{\prime}_{1}-1}
\Ts_{\emptyset}(x-k\eta) \Bigr )^{-1}
\!\! \det_{1 \le i,j \le \lambda^{\prime}_{1}}
\Ts_{\lambda_{i}-i+j}\bigl (x\! -\! (j\! -\! 1)\eta \bigr ),
\end{align}
\begin{align}\label{CBR1}
\Ts_{\lambda }(x)=\Bigl (\prod_{k=1}^{\lambda_{1}-1}
\Ts_{\emptyset}(x-k \eta ) \Bigr )^{-1}
\!\! \det_{1 \le i,j \le \lambda_{1}}
\Ts^{\lambda^{\prime}_{i}- i+j}\bigl (x\! +\! (j\! -\! 1)\eta \bigr ).
\end{align}

\subsection{The construction of the master T-operator}

The master T-operator is a generating function of the
T-operators $\Ts _{\lambda}(x)$ of a special form.
It can be introduced in the same way as in \cite{Alexandrov:2011aa}.
(In an implicit form, the notion of the master T-operator
appeared in \cite{KLT10}.) For this we should recall
the definition of the Schur functions.

Let ${\bf t}=\{t_1, t_2, t_3 , \ldots \}$ be an infinite set
of complex parameters (we call them
times) and $s_{\lambda}({\bf t})$ be the
standard Schur
functions ($S$-functions)
which can be introduced as
$$
s_{\lambda}({\bf t})=\det_{1\leq i,j\leq \lambda_1'}
h_{\lambda_i -i+j}({\bf t})\,, \qquad
$$
where the polynomials $h_k({\bf t})=s_{(k)}({\bf t})$
(the elementary Schur functions) are defined by
\begin{equation} \label{xi}
e^{\xi ({\bf t},z)}=\sum_{k=0}^{\infty}h_k({\bf t})z^k\,,
\qquad
\xi ({\bf t},z):= \sum_{n=1}^{\infty}t_k z^k.
\end{equation}
It is convenient to put $h_k({\bf t})=0$ for negative $k$
and $s_{\emptyset}({\bf t})=1$.
As is obvious from the definition, the
Schur functions are polynomials in the times $t_i$.
The Schur functions are often regarded as symmetric functions
of variables $\xi_\alpha$ such that $t_k=\frac{1}{k}\sum_\alpha
\xi_{\alpha}^k$.

The supercharacters can be expressed in terms of the
Schur functions\footnote{
This expression is equivalent to
the one in terms of the so-called supersymmetric
Schur functions ($SS$-functions)
\cite{BR87} which are symmetric functions
of two sets of variables, $\{g_a\}_{a \in \mathfrak{B}}$ and
$\{ g_{b}\}_{b \in \mathfrak{F}}$.}
as follows \cite{BB81}. Set
$y_k=\frac{1}{k}\, \mbox{str} \, \g^k$, where
$\mbox{str}\, \g ^k$ is the supertrace
of $\g ^k$ realized in the vector
representation as a $K\! \times \! K$ diagonal matrix:
$\displaystyle{\mbox{str}\,
\g ^k=\sum_{a=1}^{N+M} (-1)^{\ps(a)} g_a^k }$.
Then $\chi_{\lambda}(\g )=s_{\lambda}(\mathbf{y})$.
This is equivalent to the fact that
$\bigl ( \mbox{sdet}\, (\1 -z\g )\bigr ) ^{-1}$ is the
generating function for the supercharacters corresponding to
one-row diagrams:
$\bigl ( \mbox{sdet}\, (\1 -z\g )\bigr ) ^{-1}=
\sum_{k\geq 0}h_k (\mathbf{y})z^k$.
(For diagonal matrices 
$\mathrm{sdet}\, \g =\prod_a g_{a}^{1-2\ps (a)}$.)
For later use we need the following identity for the supercharacters:
\beq\label{Cauchy-Littlewood}
\sum_{\lambda}\chi_{\lambda}(\g ) s_{\lambda}({\bf t})=
\exp \Bigl (\sum_{k\geq 1}t_k \, \mbox{str}\, \g ^k\Bigr ),
\eeq
which is simply the
Cauchy-Littlewood identity
for the Schur functions \cite{Macdonald}.
Here and below,
the sum $\sum_{\lambda}$
goes over all Young diagrams $\lambda$ including the empty one.

Now we are ready to introduce the master T-operator
as an infinite sum over the Young diagrams:
\begin{align}
\Ts(x,{\mathbf t})=\sum_{\lambda }
\Ts_{\lambda}(x) s_{\lambda}({\mathbf t}).
 \label{masterT-co}
\end{align}
It immediately
follows from the definition
that the T-operators $\Ts _{\lambda}(x)$ can be restored from it
by applying the differential operators in the times $t_i$:
\beq\label{master3}
\left. \phantom{A_B}
\Ts _{\lambda}(x)=s_{\lambda}(\tilde \p )
\Ts(x,{\mathbf t})\right |_{{\mathbf t}=0},
\eeq
where $\tilde \p := \bigl \{ \p_{t_1}, \frac{1}{2}\, \p_{t_2},
\frac{1}{3}\, \p_{t_3}, \ldots \bigr \}$.
In particular,
$\displaystyle{\Ts _{\emptyset}(x)=\Ts (x, {\bf 0})=
\prod_{j=1}^L (x-x_j)\, {\bf I}}$
and
$\Ts_{\Box}(x)=\p_{t_1}
\Ts(x,{\mathbf t})\Bigr |_{{\mathbf t}=0}$, so that
the T-operator $\Tf (x)=\Tf _{(1)}(x)=\Tf_{\Box}(x)$
(\ref{Top0}) is expressed as the logarithmic
derivative of the master T-operator:
\beq\label{master3a}
\Tf (x)=\p_{t_1}\log \Ts (x, \tf )\Bigr |_{\tf =0}.
\eeq

Using (\ref{ntl}) and the Cauchy-Littlewood identity
(\ref{Cauchy-Littlewood}), one can derive the following
expansion of the master T-operator as $x\to \infty$:
\beq\label{ntl1}
\frac{\Ts (x, \tf )}{\Ts _{\emptyset}(x)}=
e^{\mbox{\scriptsize{str}}\, \xi (\tf , \g )}
\left (\mathbf{I}+ \frac{\eta}{x}\sum_{j=1}^L \sum_{k\geq 1}
kt_k (\g ^{(j)})^k \, + \, O(1/x^2)\right ).
\eeq
Note that the sign factor $(-1)^{\ps (b)}$ coming from
the definition
of the ${\bf P}_{\lambda}^{0j}$ (\ref{Plam}) cancels
against the one coming from the supertrace, so the two
leading terms do not actually depend on the grading
(except for the supertrace in the common factor).
More generally, using (\ref{ntl3}) and the Cauchy-Littlewood identity
one arrives, in a similar way, to the following explicit
expression for the master T-operator:
\beq
\label{ntl4}
\begin{array}{lll}
\displaystyle{\frac{\Ts (x, \tf )}{\Ts _{\emptyset}(x)}}
&=&\displaystyle{\sum_{l=0}^{L}\, \eta^l
\!\!\!  \sum_{i_1< \ldots < i_l}^L
\!\!\! \sum_{{\tiny \mbox{
$\begin{array}{c}\tiny{a_1, \ldots , a_l}\\
b_1, \ldots , b_l
\end{array}$
}}}\Bigl (
\overrightarrow{\prod_{\alpha =1}^{l}}\,
\frac{(-1)^{\ps (b_{\alpha})}
\e _{b_{\alpha}a_{\alpha}}^{(i_\alpha )}}{x-x_{i_\alpha}}
\,
\frac{\partial}{\partial \varepsilon_{\alpha}}\Bigr )
\,
e^{\mathrm{str} \, \xi  \left (\mathbf{t},
e^{\varepsilon_l \e_{a_lb_l}}
\ldots e^{\varepsilon_1 \e_{a_1b_1}}\g \right )}
\Biggm |_{\varepsilon_\alpha =0}}
\\ && \\
&=&\displaystyle{
\overrightarrow{\prod_{l =1}^{L}}\left (\mathbf{I}+
\eta \sum_{a_l, b_l}\frac{(-1)^{\ps (b_l)}
\e^{(l)}_{b_l a_l}}{x-x_l}\, \frac{\partial}{\partial \varepsilon_{l}}
\right )e^{\mathrm{str} \, \xi  \left (\mathbf{t},
e^{\varepsilon_l \e_{a_lb_l}}
\ldots e^{\varepsilon_1 \e_{a_1b_1}}\g \right )}
\Biggm |_{\varepsilon_l =0}}
\end{array}
\eeq

Given $z\in \mathbb{C}$, we will use the standard
notation ${\bf t}\pm [z^{-1}]$ for the following special
shift of the time variables:
$$
\begin{array}{l}
{\bf t}\pm [z^{-1}]:= \Bigl \{t_1 \pm z^{-1}, \, t_2 \pm \frac{1}{2}
z^{-2}, \, t_3 \pm \frac{1}{3}
z^{-3}, \, \ldots \Bigr \}\end{array}.
$$
As we shall see below,
$\Ts (x, {\bf t}\pm [z^{-1}])$ regarded as functions of
$z$ with fixed $x,{\bf t}$ play an important role.
Here we only note that equation (\ref{master3}) implies that
$\Ts (x, {\bf 0}\pm [z^{-1}])$ are the generating series for the
T-operators corresponding to the one-row and one-column diagrams
respectively:
\begin{equation}\label{master4}
\Ts (x,  [z^{-1}])=\sum_{s=0}^{\infty}z^{-s}\Ts_{s}(x), \quad \quad
\Ts (x,  -[z^{-1}])=\sum_{a=0}^{\infty}(-z)^{-a}\Ts^{a}(x).
\end{equation}

\subsection{The master T-operator and the mKP hierarchy}

\subsubsection{The bilinear identity for the master T-operator}

The main property of the master T-operator
which provides a remarkable
link to the theory of classical non-linear integrable
equations and their hierarchies is given by the following statement.
\begin{theor}
The master T-operator \eqref{masterT-co} satisfies the
bilinear identity for the mKP hierarchy:
\begin{align}
\oint_{{\mathcal C}} z^{(x-x')/\eta} e^{\xi ({\mathbf t}-{\mathbf t'},z)}
\, \Ts \bigl (x, {\mathbf t}-[z^{-1}]
\bigr )\Ts \bigl (x',{\bf t'}+[z^{-1}]\bigr )dz =0
 \label{bilinear}
\end{align}
for all ${\mathbf t}$, ${\mathbf t}'$, $x$ and $x'$.
The integration contour ${\mathcal C}$ encircles the cut
$[0, \infty ]$ between $0$ and $\infty$ and does not enclose
any singularities coming from the $\Ts$-factors.
\end{theor}

\noindent
The bilinear identity has superficially the same form as
the one for the master T-operator
for the ordinary (non-supersymmetric)
spin chains associated with $Y(gl(N))$ \cite{Alexandrov:2011aa}
and can be proved in a similar way. The proof is based
on the CBR formulas (\ref{CBR}) or (\ref{CBR1}).
The definition of the master T-operator
(\ref{masterT-co}) can be interpreted
as the expansion of the tau-function
in Schur polynomials \cite{Sato,Orlov-Shiota,EH}.
The functional relations for quantum transfer matrices
\cite{Cherednik89,BR90,KNS,Tsuboi:1997iq} are then the
Pl\"ucker-like relations for coefficients of the expansion
\footnote{It is pertinent to note possible generalizations
of this picture.
There are functional relations \cite{GKV09,Hegedus09} and
their Wronskian-like determinant
solutions \cite{Gromov:2010km,Tsuboi11}
related to infinite-dimensional representations of
$gl(N|M)$. The solutions are
given by changing the expansion point of
the generating function for the T-operators.
Therefore, there is a possibility that
the master T-operator \eqref{masterT-co} is still relevant
for such systems after a sort of analytic continuation.
In this paper, we consider only covariant tensor representations of
$gl(N|M)$.
There are also contravariant and mixed representations
whose characters are labelled by a pair of Young diagrams.
It is an interesting open problem
whether the corresponding master T-operator is a tau-function
of any hierarchy of soliton equations (like
the 2D Toda lattice).
}.

The bilinear identity
(\ref{bilinear}) is a source of various
bilinear Hirota equations for the master T-operator.
For example, setting $x'=x-\eta$,
$\tf ' =\tf -[z_1^{-1}]-[z_2^{-1}]$,
we obtain the 3-term difference Hirota equation
\begin{equation}\label{bi3}
\begin {array}{c}
z_2\Ts \left (x+\eta ,{\bf t}-[z_{2}^{-1}]\right )
\Ts \left (x,{\bf t}-[z_{1}^{-1}]\right )-
z_1 \Ts \left (x+\eta , {\bf t}-[z_{1}^{-1}]\right )
\Ts \left (x, {\bf t}-[z_{2}^{-1}]\right )
\\ \\
+\, (z_1-z_2)\Ts (x+\eta , {\bf t})\Ts
\left (x, {\bf t}-[z_{1}^{-1}]-[z_{2}^{-1}]\right )\, =0
\end{array}
\end{equation}
which is in fact equivalent to the bilinear
identity (see \cite{Shigyo}).

\subsubsection{The Baker-Akhiezer functions}

Let $T(x, \tf )$ be any eigenvalue of the master
T-operator. As it follows from (\ref{bilinear}), it is
a tau-function of the mKP hierarchy.
It is then natural to incorporate other key ingredients
of the soliton theory. The most important for us
are the Baker-Akhiezer (BA) function and
its adjoint. In what follows we refer to both as the BA functions.
They are defined as
\begin{align}
\psi (x,\tf ; z) &=z^{x/\eta} e^{\xi(\tf,z)} \,
  \frac{T (x,\tf -[z^{-1}])}{T(x,\tf)} ,
 \label{BA-fun1}
\\
\psi^{*}(x, \tf ; z) &=z^{-x/\eta} e^{-\xi(\tf,z)} \,
  \frac{T (x,\tf +[z^{-1}])}{T (x,\tf)} .
 \label{BA-fun2}
\end{align}
We are going to also consider the operator-valued BA functions
$\hat \psi (x,\tf ; z)$
defined by the same formulas with $T (x, \tf )$ substituted by
$\Ts (x, \tf )$. Since these operators commute for all $x, \tf$,
the operator $\hat \psi (x,\tf ; z)$ is well-defined.

According to the definition
of the master T-operator, $\Ts \bigl (x, \tf \mp [z^{-1}] \bigr )$
is an infinite series in $z^{-1}$. From
(\ref{ntl4}) one can see that this series converges
to a rational function of $z$ for any $x, \tf$
if $|z|>\mbox{max}\, \bigl \{|g_1|, |g_2|, \ldots , |g_{K}|
\bigr \}$. Explicitly, we obtain:
%
\beq\label{ntl5}
\begin{array}{c}
\displaystyle{
\Ts \bigl (x, \tf \mp [z^{-1}]\bigr )=
\overrightarrow{\prod_{l =1}^{L}}\left ((x-x_l)\mathbf{I}+
\eta \sum_{a_l, b_l}(-1)^{\ps (b_l)}
\e^{(l)}_{b_l a_l}\, \frac{\partial}{\partial \varepsilon_{l}}
\right )
}
\\ \\
\displaystyle{\times \,\,\,
\!\! \Biggl \{ \Bigl [ \mathrm{sdet}\, \bigl ( \1 \! - \! z^{-1}
\g_{\, \varepsilon_L,\, \ldots , \, \varepsilon_1}^{a_Lb_L,
\ldots , a_1b_1} \bigr ) \Bigr ]^{\pm 1}
e^{\mathrm{str} \, \xi  \left (\mathbf{t},
\g_{\, \varepsilon_L,\, \ldots , \, \varepsilon_1}^{a_Lb_L,
\ldots , a_1b_1} \right )}\Biggr \}\Biggm |_{\varepsilon_l =0}},
\end{array}
\eeq
where we have put
$\g_{\, \varepsilon_n,\, \ldots , \, \varepsilon_1}^{a_nb_n,
\ldots , a_1b_1}\equiv
e^{\varepsilon_n \e_{a_nb_n}}\ldots
e^{\varepsilon_1 \e_{a_1b_1}}\g$ for brevity.
Therefore, the function
$z^{-x/\eta} e^{-\xi(\tf ,z)}  \psi (x,\tf ; z)$
(resp.\ $z^{x/\eta} e^{\xi(\tf ,z)} \psi^{*}(x, \tf ; z) $)
is a rational function of $z$ with poles at the points
$z=g_{a}$ (the eigenvalues of the matrix $\g$)
for $a \in \mathfrak{F} $ (resp.,\  $a \in \mathfrak{B} $ )
of at least first order.
It can be also derived\footnote{
The
main underlying statement is that
$$
\begin{array}{c}
\displaystyle{
\left (x\mathbf{I}+\eta \sum_{a,b}(-1)^{\ps (b)}\e_{ba}^{(l)}
\frac{\p}{\p \varepsilon}\right ) \bigl [\mbox{sdet}  \bigl (
e^{\varepsilon \e_{ab}}\g \bigr )\bigr ]^{\pm 1} \, \Phi (\varepsilon )
\Bigm |_{\varepsilon =0}}
\\ \\
\displaystyle{=\,\, \bigl [\mbox{sdet} \g \bigr ]^{\pm 1}
\left ((x\pm \eta )\mathbf{I}+\eta \sum_{a,b}(-1)^{\ps (b)}\e_{ba}^{(l)}
\frac{\p}{\p \varepsilon}\right )\, \Phi (\varepsilon )
\Bigm |_{\varepsilon =0}}
\end{array}
$$
for any $\g \in GL(M|N)$ and any $\Phi \in \mbox{End}
\bigl (V_{l+1}\otimes V_{l+2} \otimes \ldots \otimes V_{L}\bigr )$
(here $V_i \cong \mathbb{C}^{N|M}$).
It immediately follows from the Leibniz rule and the
identity
$$
\frac{\p}{\p \varepsilon}
\bigl [\mbox{sdet}  \bigl (
e^{\varepsilon \e_{ab}}\g \bigr )\bigr ]^{\pm 1}
\Bigm |_{\varepsilon =0}=\mbox{str}( \e_{ab} )
\bigl [\mbox{sdet} \g \bigr ]^{\pm 1}=
(-1)^{\ps (b)}\delta_{ab}\bigl [\mbox{sdet} \g \bigr ]^{\pm 1}
$$
which is easy to check.
}
from (\ref{ntl5}) that
\begin{align}
\lim_{z \to 0} z^{\pm(N-M)}\Ts (x,\mathbf{t} \mp [z^{-1}])&=
(-1)^{N-M}(\mathrm{sdet}\, \g )^{\pm 1}
\Ts (x \pm \eta ,\mathbf{t}).
 \label{limT}
\end{align}
The left hand side is to be understood as the analytic
continuation to the point $z=0$ of the analytic
function (rational in our case) defined by the
series which converges in a neighbourhood of infinity.

Since any eigenvalue $T(x,\tf )$ is a polynomial in $x$, the
functions $z^{-x/\eta}\psi$ and
$z^{x/\eta}\psi^*$,
regarded as functions of $x$, are rational
functions with $L$ zeros and
$L$ poles which are simple in general position.
From \eqref{masterT-co}, \eqref{Top-poly} and \eqref{Top},
using the Cauchy-Littlewood identity, or directly
from (\ref{ntl5}), we conclude that
$$
\lim_{x\to \infty}
\frac{T(x, \tf \mp [z^{-1}])}{T(x, \tf )} =
\frac{e^{\mathrm{str} \, \xi (\tf \mp [z^{-1}], \g )}}{e^{\mathrm{str}
\, \xi (\tf , \g )}}=
e^{\mathrm{str} \, \xi (\mp [z^{-1}], \g )}=
\left [ \mathrm{sdet} \left ( \1 \! -\!
z^{-1}\g \right )\right ]^{\pm 1},
$$
hence
\begin{align}
&\lim_{x \to \infty}
 z^{-x/\eta} e^{-\xi(\tf ,z)}  \psi (x, \tf;, z) =
 z^{-N+M} \mathrm{sdet} (z\1 -\g ),
 \label{psi-lim}
\\
& \lim_{x \to \infty}
z^{x/\eta} e^{\xi(\tf ,z)}  \psi^{*}(x, \tf ; z) =
 z^{N-M} \bigl ( \mathrm{sdet} (z\1 -\g) \bigr )^{-1} .
\end{align}

The (operator-valued) functions
$\hat \psi (x, z):=\hat \psi (x, \mathbf{0}; z)$ and
$\hat \psi^{*}(x, z):=\hat \psi^{*}(x, \mathbf{0}; z)$,
as well as the corresponding eigenvalues, are called
{\it stationary} BA functions.
Their explicit form directly follows from (\ref{ntl5}):
{\small
\beq\label{ntl6}
z^{-x/\eta}\hat \psi (x,z) \! =\!
\sum_{l=0}^{L} \eta^l
\!\!\!  \sum_{i_1 < \ldots < i_l}^L
\!\!\! \sum_{{\tiny \mbox{
$\begin{array}{c}\tiny{a_1, \ldots , a_l}\\
b_1, \ldots , b_l
\end{array}$}}}\Bigl (
\overrightarrow{\prod_{\alpha =1}^{l}}\,
\frac{(-1)^{\ps (b_{\alpha})}
\e _{b_{\alpha}a_{\alpha}}^{(i_\alpha )}}{x-x_{i_\alpha}}
\,
\frac{\partial}{\partial \varepsilon_{\alpha}}\Bigr )
\,
\mathrm{sdet}\,
\bigl ( \1 \! - \! z^{-1}
\g_{\, \varepsilon_l,\, \ldots , \, \varepsilon_1}^{a_lb_l,
\ldots , a_1b_1}\bigr )
\! \Biggm |_{\varepsilon_\alpha =0},
\eeq
\beq\label{ntl6a}
z^{x/\eta}\hat \psi^* (x,z) \! =\!
\sum_{l=0}^{L} \eta^l
\!\!\!  \sum_{i_1 <\ldots < i_l}
\!\!\!\! \sum_{{\tiny \mbox{
$\begin{array}{c}\tiny{a_1, \ldots , a_l}\\
b_1, \ldots , b_l
\end{array}$}}}\Bigl (
\overrightarrow{\prod_{\alpha =1}^{l}}\,
\frac{(-1)^{\ps (b_{\alpha})}
\e _{b_{\alpha}a_{\alpha}}^{(i_\alpha )}}{x-x_{i_\alpha}}
\,
\frac{\partial}{\partial \varepsilon_{\alpha}}\Bigr )
\,
 \bigl [ \mathrm{sdet}\, \bigl ( \1 \! - \! z^{-1}
\g_{\, \varepsilon_l,\, \ldots , \, \varepsilon_1}^{a_lb_l,
\ldots , a_1b_1}\bigr )
\bigr ]^{-1}
\! \Biggm |_{\varepsilon_\alpha =0}
.
\eeq
}
In particular, we have the expansion of
$\hat \psi (x,z)$ as $|x|\to \infty$:
\beq\label{ntl2}
z^{-x/\eta}\hat \psi (x,z)=
\mbox{sdet} \, \bigl (\1 \!
-\! z^{-1}\g \bigr )
\left (\mathbf{I} \, - \, \frac{\eta}{x}\sum_{j=1}^L\sum_{a=1}^{K}
\frac{g_a \e_{aa}^{(j)}}{z-g_a}\,  \, +\, O(1/x^2) \right )
\eeq
(we need it in the next section).

For calculations in the next section we also need the following
general properties of the BA functions of the mKP hiertarchy.
\begin{itemize}
\item[a)]
They obey the differential-difference
equations of the form
\beq\label{dif1}
\partial_{t_{1}} \psi (x, \tf ; z) =
 \psi(x\! +\! \eta ,\tf ; z)
 +V(x,\tf )\, \psi (x, \tf; z),
\eeq
\beq\label{dif2}
-\partial_{t_{1}} \psi^{*}(x, \tf ; z) =
  \psi^{*}(x\! -\! \eta , \tf ; z)
 +V(x\! -\! \eta ,\tf )\psi^{*}(x, \tf ; z)
\eeq
(the linear problems), where
$\displaystyle{
V(x,\tf )=
\partial_{t_{1}} \log \frac{T(x\! +\! \eta , \tf )}{T(x, \tf )}}
$
(see, e.g., \cite{DJKM83,JM83});
\item[b)]
They obey the relation
\begin{align}\label{dif3}
\partial_{t_{m}}
\log \frac{T(x\! +\! \eta ,\mathbf{t})}{T(x,\mathbf{t})}&=
\mathrm{res}_{\infty}
\bigl (\psi (x, \tf ; z) \psi^{*}(x\! +\! \eta ,
\tf ; z \bigr ) z^{m}\mathrm{d}z),
\end{align}
where the residue is normalized as
$\mathrm{res}_{\infty}\, z^{-1}\mathrm{d}z =1$.
It can be derived from
\eqref{bilinear} and \eqref{limT} in the same way as
in \cite{Z13}.
\end{itemize}

\section{From the master T-operator to the
classical RS model and back}

\subsection{Eigenvalues of the spin chain Hamiltonians
as velocities of the RS particles}

As we already mentioned in the previous section,
the eigenvalues of the master T-operator are polynomials in the
spectral parameter $x$ of degree $L$:
\begin{align}\label{tau1}
T(x, \tf)=e^{\, \mathrm{str} \, \xi (\tf , \g )}
\prod_{k=1}^{L}\bigl (x-x_{k}(\tf)\bigr ).
\end{align}
The roots have their own dynamics in the times.
The very fact that $T(x, \tf )$ is a tau-function of the
mKP hierarchy implies \cite{KZ95,Iliev} that
the roots $x_i$
move in the time $t_1$ as particles of the RS $L$-body system
\cite{RS}. Moreover, their motion in the higher times $t_k$
is the same as motion of the RS particles caused by the higher
Hamiltonian flows of the RS system
(see \cite{Iliev,Z13} which extend the methods
developed by Krichever \cite{Krichever-rat} and Shiota \cite{Shiota}).
We have
$\displaystyle{T(x,\mathbf{0})=
T_{\emptyset}(x)=\prod_{k=1}^{L}(x-x_{k})}$,
where $x_{k}=x_{k}(\mathbf{0})$. This means that
\begin{itemize}
\item[{\rm (i)}]
The inhomogeneity parameters $x_k$ of the spin chain
should be identified with {\it initial positions} $x_k(\mathbf{0})$
of the RS particles.
\end{itemize}
With the help of (\ref{master3a}) we can write:
\begin{align}
\frac{T_{\Box}(x)}{T_{\emptyset}(x)}=
\partial_{t_{1}} \log T(x, \tf)|_{\tf =0}=
\mathrm{str} \, \g
 -\sum_{k=1}^{L}
 \frac{\dot{x}_{k}(\mathbf{0})}{x-x_{k}},
\end{align}
where $ \dot{x}_{k}(\mathbf{0}):=
\partial_{t_{1}}x_{k}(\tf)|_{\tf=0}$.
Comparing this with \eqref{T1ex}, we find:
\begin{align}\label{xdotH}
\dot{x}_{k}(\mathbf{0})=-\eta H_{k},
\end{align}
where $H_k$ is an eigenvalue of $\mathbf{H}_{k}$.
Therefore, in addition to (i) we conclude that
\begin{itemize}
\item[{\rm (ii)}]
The eigenvalues $H_k$ of the susy-XXX spin chain Hamiltonians
are expressed through the {\it initial
velocities} of the RS particles as
$H_i=-\dot x_i (\mathbf{0})/\eta$.
\end{itemize}
In other words, any point in the phase space of the $L$-body RS
system with coordinates $\{x_i, \dot x_i\}$
corresponds to an eigenstate,
with the eigenvalues $H_i=-\dot x_i/\eta$, of the Hamiltonians of the
susy-XXX spin chain on $L$ sites with the inhomogeneity parameters
$x_i$.

This unexpected connection between quantum spin chains
and the classical RS model was pointed out in \cite{Alexandrov:2011aa}
as a corollary of the Hirota bilinear equations for the master
T-operator.
A similar relation between quantum Hamiltonians
in the Gaudin model and velocities of particles in the classical
Calogero-Moser model was found in \cite{MTV} using different
methods (see also \cite{MTV1,MTV2}
for further developments)\footnote{
In \cite{Alexandrov:2013aa} it was obtained from the master
T-operator construction for the Gaudin model.}
The message of the present paper is that the identifications
(i) and (ii) are in fact independent of the grading: their
form is the same for
all spin chains of the XXX type associated with {\it any}
(super)algebra $gl(N|M)$ including the ordinary algebras
$gl(N|0)=gl (N)$.

\subsection{Lax pair for the RS model from
dynamics of poles}

To make the correspondence ``quantum spin chains $\leftrightarrow$
classical RS systems''
(the QC correspondence) complete, we need the Lax matrix
for the RS model.
At this stage the set-up is exactly the same as in \cite{Z13}.
Here we repeat
the main formulas with some comments skipping the details.

Below we will derive equations of motion
for the $t_1$-dynamics of the $x_i$'s using Kri\-che\-ver's method
\cite{Krichever-rat}, the starting point of which is
the linear problem (\ref{dif1}) for the BA function.
Essentially, the derivation is
not specific to the master $T$-operator case but only depends on
the polynomial form of the tau-function.

One can derive equations of motion for the $x_i$'s performing
the pole expansion of the linear problem (\ref{dif1}).
It is convenient to
denote $t_{1}=t$ and put all higher times equal to $0$ because
they are irrelevant for this derivation.
Below in this section we often write simply $t$ instead of $\tf$.
According to \eqref{BA-fun1}, the general form of
$\psi$ as a function of $x$ is
\begin{align}
\psi (x,t; z)=z^{x/\eta}e^{tz}
 \left (c_{0}(z)+ \sum_{j=1}^{L}\frac{c_{j}(z,t)}{x-x_{j}(t)} \right),
\end{align}
where $ c_{0}(z)=\mathrm{sdet} (\1 -z^{-1}\g)$
(see \eqref{psi-lim}).
One should substitute it into the linear equation
(\ref{dif1}) with
\beq\label{CM3}
V(x,t )=\p_t \log \frac{T(x+\eta ,t )}{T(x,t )}
=\sum_{k=1}^{L}\left (\frac{\dot x_k}{x-x_k}-
\frac{\dot x_k}{x-x_k+\eta}\right ), \quad x_k=x_k(t)
\eeq
and cancel all the poles at $x=x_i$ and $x=x_i-\eta$
(possible poles of the second order cancel automatically).
This yields an overdetermined system of linear equations for
the coefficients $c_i$:
\beq\label{CM5}
\left \{
\begin{array}{l}
(z\1 -\mathsf{Z})\, \vec{c}=c_0 (z) \, {\sf \dot X}\, \vec{1}
\\ \\
\dot{\vec {c}}=
\mathsf{G}\, \vec{c},
\end{array}
\right.
\eeq
where $\vec {c}=(c_1, c_2, \ldots , c_L)^{\sf t}$,
$\vec {1} =(1, 1, \ldots , 1)^{\sf t}$ are $L$-component
vectors and the $L \! \times \!L$ matrices
$\mathsf{X}=\mathsf{X}(t)$,
$\mathsf{Z}=\mathsf{Z}(t)$, $\mathsf{G}=\mathsf{G}(t)$ are
defined by their matrix elements as follows:
\beq\label{CM6}
\mathsf{X}_{ij}=x_i \delta_{ij}, \quad \quad
\mathsf{Z}_{ij}= \frac{\dot x_i}{x_i -x_j -\eta}
\eeq
\beq\label{CM7}
\mathsf{G}_{ij}=
\left (
\sum_{k\neq i}\frac{\dot x_k}{x_i -x_k} \! -
\!\sum_{k\neq i} \frac{\dot x_k}{x_i \! -\! x_k \! +\! \eta}
\right )\delta_{ij}
+\left (\frac{\dot x_i}{x_i \! -\! x_j}-
\frac{\dot x_i}{x_i \! -\! x_j\!  -\! \eta}\right )
\bigl (1\! -\! \delta_{ij}\bigr ).
\eeq
The explicit form of the matrix $\mathsf{G}$ is not used
in what follows.
As is easy to check, the matrix $[\mathsf{X}, \, \mathsf{Z}]-
\mathsf{Z}$ has rank $1$. More precisely, these matrices satisfy the
commutation relation
\beq\label{comm}
[\mathsf{X}, \, \mathsf{Z}]=\eta \mathsf{Z}
+\mathsf{\dot X}\mathsf{E}\,,
\eeq
where $\mathsf{E}=\vec{1}
\otimes \vec{1}^{\sf t}$ is the $L\! \times \! L$
matrix of rank $1$ with all entries equal to $1$.
As a consequence of this commutation relation, we mention the identity
$\vec{1}^{\sf t}\mathsf{Z}^k \mathsf{\dot X}\, \vec 1=
-\eta \, \mbox{tr}\, \mathsf{Z}^{k+1}$
which holds for any $k\geq 0$
(see \cite{Z13}).

The compatibility condition of the problems (\ref{CM5}) is the Lax
equation
\beq\label{CM8}
\mathsf{\dot Z}=[\mathsf{G}, \, \mathsf{Z}]
\eeq
which is equivalent to the equations of motion
\beq\label{CM9}
\ddot x_i = -\sum_{k\neq i}
\frac{2\eta^2 \dot x_i \dot x_k}{\bigl (x_i \! -\! x_k\bigr )
\bigl [(x_i\! -\! x_k)^2-\eta^2\bigr ]}\,, \quad \quad i=1, \ldots , L.
\eeq
This dynamical system called the RS model
is sometimes referred to as the relativistic
deformation of the Calogero-Moser model,
the parameter $\eta$ being the inverse ``velocity of light''.
The Hamiltonian formulation is given in the appendix.
The integrability of the RS model follows from the Lax representation.
The matrix $\mathsf{Z}$ is the Lax matrix.
As it follows from (\ref{CM8}), the time evolution preserves
its spectrum, i.e., the coefficients
${\cal J}_k$ of the characteristic polynomial
\beq\label{char-pol}
\det (z\1 -\mathsf{Z}(t))=\sum_{k=0}^{L}{\cal J}_k z^{L-k}
\eeq
are integrals of motion. Equivalently, one can say that
eigenvalues of the Lax matrix are integrals of motion.

For completeness, we also present here the linear problem
for coefficients of the adjoint
BA function
\begin{align}
\psi^* (x,t; z)=z^{-x/\eta}e^{-tz}
 \left (c_{0}^{-1}(z)+ \sum_{j=1}^{L}\frac{c^*_{j}(z,t)}{x-x_{j}(t)}
 \right).
\end{align}
As a counterpart of (\ref{CM5}), we get, using the equations of
motion,
\beq\label{CM5a}
\left \{
\begin{array}{l}
\vec{c}^{*{\sf t}}{\sf \dot X}^{-1}
(z\1 -\mathsf{Z})=-c_0^{-1} (z)\vec 1^{\sf t}
\\ \\
\p_t({\vec c}^{*{\sf t}}\mathsf{\dot X^{-1}})
=-\vec{c}^{*{\sf t}}\mathsf{\dot X}^{-1}\mathsf{G}. \end{array}
\right.
\eeq
Here $\vec{c}^{*{\sf t}}=(c_1^*, c_2^*, \ldots , c_L^*)$ and
$\vec{1}^{\sf t}=(1, 1, \ldots , 1)$ (note that
$\vec{1}^{\sf t}\mathsf{G}=0$). Regarding these equations as
overdetermined linear problems for the (co)vector
${\vec c}^{*{\sf t}}\mathsf{\dot X^{-1}}$, one comes to the
same Lax equation (\ref{CM8}) as their compatibility condition.
The adjoint linear problems (\ref{CM5a}), together with
general relation (\ref{dif3}), are used for the extension
of the time dynamics of the $x_i$'s to the whole hierarchy,
as it has been done in \cite{Iliev,Z13}, see Appendix B.

\subsection{The BA function and the master T-operator}

The solution for the vector $\vec c$ reads
$\vec c(z,t)=c_0(z)\bigl (z\1 -\mathsf{Z}(t)\bigr )^{-1}\,
\mathsf{\dot X}\, \vec 1$.
The BA function $\psi$ is then given by the formula
\beq\label{cc1}
\psi =c_0(z)\, z^{x/\eta} e^{tz}\Bigl (
1+\vec{1}^{\sf t}(x\1 -\mathsf{X})^{-1}
(z\1 -\mathsf{Z})^{-1}\mathsf{\dot X}{\vec 1}\Bigr ).
\eeq
Similar formulas can be obtained for the
adjoint vector $\vec{c}^{*{\sf t}}$ and the adjoint BA function:
$$
\vec{c}^{*{\sf t}}(z,t)=-\, c_0^{-1}(z)\, \vec{1}^{\sf t}\, (z\1 \! -\!
\mathsf{Z}(t))^{-1}\mathsf{\dot X},
$$
\beq\label{cc1a}
\psi^* =c_0^{-1}(z)\, z^{-x/\eta} e^{-tz}\Bigl (
1-\vec{1}^{\sf t}(z\1 -\mathsf{Z})^{-1}
(x\1 -\mathsf{X})^{-1}\mathsf{\dot X}{\vec 1}\Bigr ).
\eeq
It is easy to see that for non-zero values of the higher times
the BA functions are given by the same formulas with the factor
$e^{\pm tz}$ substituted by $e^{\pm \xi (\tf , z)}$.
Writing $\vec{1}^{\sf t} \mathsf{A} {\vec 1} = \mbox{tr} (
\mathsf{A}\mathsf{E})$ and using
the commutation relation (\ref{comm}),
one can represent these expressions as
ratios of determinants\footnote{The order of the factors
$x\1\! -\! \mathsf{X}$ and $z\1 \! -\!\mathsf{Z}$
under the determinant upstairs is actually not important
because of the identity $\det (\mathsf{A}\mathsf{B}+\mathsf{A}_1)=
\det (\mathsf{B}\mathsf{A}+\mathsf{A}_1)$ valid for any matrices
$\mathsf{A}, \mathsf{B}, \mathsf{A}_1$ such that
$[\mathsf{A}, \mathsf{A}_1]=0$.}:
\beq\label{BA1a}
\psi (x, \tf ;z)
=\mathrm{sdet} \bigl (\1 - z^{-1}\g\bigr ) z^{x/\eta} \,
e^{\xi (\tf , z)}\,
\frac{\det \bigl [(x\1\! -\! \mathsf{X}) (z\1 \! -\!\mathsf{Z})\! -
\! \eta \mathsf{Z}\bigr ]}{\det (x\1-\mathsf{X})\det (z\1-\mathsf{Z} )},
\eeq
\beq\label{BA1b}
\psi^* (x, \tf ;z)
=\Bigl [\mathrm{sdet} \bigl (\1 \! -\!  z^{-1}\g\bigr )\Bigr ]^{-1}
z^{-x/\eta} \,
e^{-\xi (\tf , z)}\,
\frac{\det \bigl [(z\1 \! -\!\mathsf{Z})(x\1\! -\! \mathsf{X}) \! +
\! \eta \mathsf{Z}\bigr ]}{\det (x\1-\mathsf{X})\det (z\1-\mathsf{Z} )}.
\eeq
In particular, the stationary BA functions are given by
\beq\label{BA1st}
\psi (x, z)
=\mathrm{sdet} \bigl (\1 - z^{-1}\g\bigr ) z^{x/\eta} \,
\frac{\det \bigl [(x\1\! -\! \mathsf{X}_0) (z\1 \! -\!\mathsf{Z}_0)\! -
\! \eta \mathsf{Z}_0\bigr ]}{\det (x\1-\mathsf{X}_0)
\det (z\1-\mathsf{Z}_0)},
\eeq
\beq\label{BA1sta}
\psi^* (x, z)
=\Bigl [\mathrm{sdet} \bigl (\1 - z^{-1}\g\bigr )\Bigr ]^{-1}
z^{-x/\eta} \,
\frac{\det \bigl [(z\1 \! -\!\mathsf{Z}_0)(x\1\! -\! \mathsf{X}_0) \! +
\! \eta \mathsf{Z}_0\bigr ]}{\det (x\1-\mathsf{X}_0)
\det (z\1-\mathsf{Z}_0)}.
\eeq
Hereafter, we use the notation $\mathsf{X}_0 =\mathsf{X}(\mathbf{0})$,
$\mathsf{Z}_0 =\mathsf{Z}(\mathbf{0})$.
Using
(\ref{BA-fun1}), one can obtain from (\ref{BA1a}) an explicit
determinant formula for eigenvalues of the master $T$-operator:
\beq\label{CM13}
T (x, {\bf t})=e^{{\scriptsize{\mbox{str}}\, \xi ({\bf t}, \g)}}
\det \Bigl ( x\1 -\mathsf{X}_0 +
\sum_{k\geq 1}kt_k \mathsf{Z}_0^{k}\Bigr ).
\eeq
Formulas (\ref{BA1a}) and (\ref{CM13}) are not new
in the context
of classical integrable hierarchies
(see, e.g., \cite{Shiota,Wilson,Iliev}). The new observation is
the close connection with quantum spin chains.
It is important to stress that we can understand
(\ref{BA1st}) and (\ref{CM13}) in the operator sense,
i.e. as expressions for the {\it quantum operators}
$\hat \psi (x, z)$, $\Ts (x, \tf )$ in terms of the
matrices $(\mathsf{X}_0)_{ij}=x_i
\delta_{ij}\mathbf{I}$,
$\displaystyle{(\mathsf{Z}_0)_{ij}=\frac{\eta \mathbf{H}_i}{x_j-x_i+\eta}}$.
The latter is
the Lax matrix $\mathsf{Z}_0$ with
the operator-valued entries given by equation (\ref{CM6})
with the substitution (\ref{xdotH}).

To summarize, we have derived equations of motion
of the RS model for roots
of the master T-operator, together with the Lax representation.
Then, embedding the initial quantum problem into the context
of the classical RS model, we have obtained explicit operator
expressions for the BA function and the master T-operator in terms
of the quantum spin chain Hamiltonians. In the next section we
show how one can reformulate the spectral problem for the
quantum Hamiltonians in terms of an inverse spectral problem
for the RS Lax matrix.

\section{Spectrum of the spin chain Hamiltonians from the
classical RS model}

\subsection{Twist parameters as eigenvalues of the Lax matrix}

The expansion of the
stationary operator-valued BA function $\hat \psi$ at large $|x|$
has the same form as in the
$gl(N)$ case \cite{Z13} except
for the overall super-determinant.
In fact we have two different (but equivalent) expressions
for $\hat \psi$: (\ref{ntl6}) and (\ref{BA1st}).
Let us compare their large $|x|$ expansions.
The large $|x|$ expansion of (\ref{ntl6}) is given by (\ref{ntl2}):
$$
\hat \psi (x,z)=c_0(z) z^{x/\eta}
\left (\mathbf{I} \, - \, \frac{\eta}{x}\sum_{j=1}^L\sum_{a=1}^{K}
\frac{g_a \e_{aa}^{(j)}}{z-g_a}\,  \, +\, O(x^{-2}) \right ),
$$
where $g_a$ are the twist parameters.
The expansion of (\ref{BA1st}) is
$$
\hat \psi (x,z)=c_0(z) z^{x/\eta} \left (
1-\frac{\eta}{x}\, \mbox{tr} \,\frac{\mathsf{Z}_0}{z\1-
\mathsf{Z}_0} \, +O(x^{-2})\right ).
$$
Equating the $O(1/x)$ terms of the two expansions leads to
the relation
\beq\label{ccc1}
\mbox{tr} \,\frac{\mathsf{Z}_0}{z\1-\mathsf{Z}_0} =
\sum_i \sum_a \frac{\e_{aa}^{(i)}g_a}{z-g_a}
\eeq
which has to be valid identically.
Let us stress that its left hand side is well-defined because
the entries of the matrix $\mathsf{Z}_0$ are commuting operators.
Using the identity $\mbox{tr}\, (z\1-\mathsf{A})^{-1}=
\p_z \log \det (z\1-\mathsf{A})$ valid for any matrix
$\mathsf{A}$, we integrate (\ref{ccc1}) to obtain
\beq\label{cc2}
\det \bigl (z\1-\mathsf{Z}_0\bigr )
=\prod_{a=1}^{K}(z-g_a)^{\mathbf{M}_a},
\eeq
where $\mathbf{M}_a$ are the weight operators(\ref{Ma}).
Since the time evolution is an isospectral deformation,
the same is true for the Lax matrix $\mathsf{Z}(\tf )$ for any
values of the times. We see that $\mathbf{M}_a$ is
the ``operator multiplicity" of the
eigenvalue $g_a$.
In the weight space ${\cal V}(\{M_a\})$ the
multiplicities become equal to the $M_a$'s.
The conclusion is:
\begin{itemize}
\item
The Lax matrix $\mathsf{Z}$ has
eigenvalues $g_a$ with multiplicities $M_a\geq 0$ such that
$M_1 +\ldots +M_N =L$.
\end{itemize}
Our next goal is to formulate the
QC correspondence\footnote{
The QC correspondence can be traced back to \cite{GK}, where joint spectra
of some commuting
finite-dimensional operators were linked to the classical
Toda chain.} between the spin chains and the RS model.

\subsection{The QC correspondence}

Consider the Lax matrix $\mathsf{Z}_0$ of the
$L$-particle RS model, where
the inverse ``velocity of light'', $\eta$, is identified with the
parameter $\eta$ introduced in the quantum $R$-matrix (\ref{R1}),
and the initial coordinates and velocities of the particles
are identified, respectively,
with the inhomogeneity parameters $x_i$
and eigenvalues of the Hamiltonians $H_i$ through $\dot x_i=-\eta H_i$:
\beq\label{QC1}
{\sf Z}_0
=\left ( \begin{array}{ccccc}
\displaystyle{\mathit{H}_1} &
\displaystyle{\frac{\eta \mathit{H}_1}{x_2\! -\!x_1\!+\! \eta}}
&\displaystyle{\frac{\eta H_1}{x_3\! -\! x_1\! +\! \eta}} &
\ldots & \displaystyle{\frac{\eta \mathit{H}_1}{x_L\! -\! x_1\! +\! \eta}}
\\ &&&& \\
 \displaystyle{\frac{\eta \mathit{H}_2}{x_1\! -\! x_2\! +\! \eta}} &
 \displaystyle{\mathit{H}_2}&
 \displaystyle{\frac{\eta \mathit{H}_2}{x_3\! -\! x_2\! +\! \eta}} &
 \ldots & \displaystyle{\frac{\eta
 \mathit{H}_2}{x_L\! -\! x_2\! +\! \eta}}
 \\ &&&& \\ \vdots & \vdots & \vdots & \ddots & \vdots
 \\ &&&& \\
 \displaystyle{\frac{\eta \mathit{H}_L}{x_1\! -\! x_L\! +\! \eta}} &
 \displaystyle{\frac{\eta \mathit{H}_L}{x_2\! -\! x_L\! +\! \eta}}&
 \displaystyle{\frac{\eta
 \mathit{H}_L}{x_3\! -\! x_L\! +\! \eta}} & \ldots &
 \displaystyle{\mathit{H}_L}
\end{array}\right ).
\eeq
According to the conclusion of the previous subsection,
we claim that if the $H_i$'s are eigenvalues of the Hamiltonians
of the spin chain in the weight space
${\cal V}(\{M_a\})\subset \mathcal{V}$, then
\beq\label{QC2}
\mbox{Spec}\, ({\sf Z}_0)=\Bigl (\underbrace{g_1,
\ldots , g_1}_{M_1}, \,
\underbrace{g_2, \ldots , g_2}_{M_2}, \, \ldots \,
\underbrace{g_{K}, \ldots , g_{K}}_{M_{K}}
\Bigr ).
\eeq
Equivalently, let $\mathcal{H}_j = \mbox{tr}\, (\mathsf{Z}_0)^j$
be the higher integrals of motion of the RS model (see Appendix B), then
their level set is defined by
$\displaystyle{{\cal H}_j=\sum_{a=1}^{K}M_a g_a^j}$.
In general, the matrix ${\sf Z}_0$ with multiple eigenvalues
is not diagonalizable and contains Jordan cells.

One can also say that the eigenstates
of the quantum Hamiltonians
correspond to the intersection points of two Lagrangian submanifolds
in the phase space of the RS model. One of them is the hyperplane
defined by fixing all the coordinates $x_i$ while the other one is
the Lagrangian submanifold obtained by fixing values
of the $L$ independent integrals of motion in involution
${\cal H}_k$, $k=1, \ldots , L$.
In general, there are many intersection points numbered
by a finite set $\mathfrak{I}$, with coordinates,
say $(x_1, \ldots , x_L, \,
p_1^{(\alpha )}, \ldots , p_L^{(\alpha )})$, $\alpha \in \mathfrak{I}$.
The values of $p_j^{(\alpha )}$ give, through equation (\ref{RS2a}),
the spectrum of ${\bf H}_j$:
$$
H_j^{(\alpha )} =e^{-\eta p_j^{(\alpha )}}\prod_{k=1, \neq j}
\frac{x_j-x_k +\eta}{x_j-x_k}\,.
$$
However, we can not claim that all the intersection points
correspond to the energy levels of the Hamiltonians for
a given spin chain.
The examples elaborated below in Appendix C suggest that there are
intersection points that do not correspond to energy levels of
a particular spin chain with a fixed grading. Instead, they
correspond to spectra of the Hamiltonians for spin chains with other
possible gradings.

Summarizing, we claim that
the spectral problem for the non-local inhomogeneous
susy-XXX spin chain
Hamiltonians ${\bf H}_j$ in the subspace ${\cal V}(\{M_a\})$
is closely linked to the following {\it inverse spectral problem}
for the RS Lax matrix ${\sf Z}_0$
of the form (\ref{QC1}).
Let us fix the spectrum of the matrix ${\sf Z}_0$ to be
(\ref{QC2}),
where $g_1, \ldots , g_{K}$ are eigenvalues
of the (diagonal) twist matrix $\g$.
Then we ask what is the set of possible
values of the $H_j$'s allowed by these constraints.
The eigenvalues $H_j$ of the quantum
Hamiltonians are contained in this set.

\subsection{Algebraic equations for the spectrum}

The characteristic polynomial of the matrix
(\ref{QC1}) can be found explicitly using
the simple fact from the linear algebra that
the coefficient in front of $z^{L-k}$
in the polynomial $\det_{L\times L} (z\1+{\sf A})$ equals
the sum of all diagonal $k\! \times \! k$ minors of the matrix ${\sf A}$.
All such minors can be found using the
decomposition $\mathsf{Z}_0= -\, \mathsf{H} \mathsf{Q}$, where
$\mathsf{H}=\mathrm{diag}\, \bigl (H_1, H_2, \ldots , H_{L}\bigr )$
and
\beq\label{Q}
\mathsf{Q}_{ij}=\frac{\eta}{x_i-x_j-\eta}\,
\eeq
is the Cauchy matrix,
and the explicit expression for the
determinant of the Cauchy matrix:
$$
\det_{1\leq i,j \leq n}\, \frac{\eta}{x_i\! -\! x_j\! -\! \eta}=
(-1)^n \prod_{1\leq i<j \leq n}
\left (1-\frac{\eta^2}{(x_i\! -\! x_j)^2}\right )^{-1}.
$$
The result is:
\beq\label{QC4}
\det_{L\times L}(z\1 -{\sf Z}_{0})=
\det_{L\times L}(z\1 +{\sf H}{\sf Q})=
\sum_{n=0}^L {\cal J}_n \, z^{L-n},
\eeq
where
\beq\label{QC5}
{\cal J}_n = (-1)^n \!\!
\sum_{1\leq i_1<\ldots < i_n \leq L}
H_{i_1}\ldots H_{i_n}
\prod_{1\leq \alpha < \beta \leq n}
\left (1-\frac{\eta^2}{(x_{i_{\alpha}}\! -\! x_{i_{\beta}})^2}\right )^{-1}.
\eeq
In particular, the highest coefficient is given by the following
simple formula:
$$
{\cal J}_L=(-1)^L \, H_1 H_2 \ldots H_L \prod_{1 \le i<j \le L}
\left (1-\frac{\eta^2}{(x_{i}\! -\! x_{j})^2}\right )^{-1}.
$$
Let us point out that the integrals ${\cal H}_k$ introduced
in the previous section are connected with the integrals ${\cal J}_k$
by the Newton's formula \cite{Macdonald}:
$\displaystyle{\sum_{k=0}^L {\cal J}_{L-k}{\cal H}_k=0}$ (we have set
${\cal H}_0=\mbox{tr} ({\sf Z})^0=L$).

Combining (\ref{QC2}) and (\ref{QC5}), we see that
the eigenvalues $H_i$ of the inhomogeneous
susy-XXX Hamiltonians can be found from the system of
polynomial equations
\begin{align}
\sum_{1 \le i_{1} < \ldots < i_{n} \le L}
 H_{i_{1}}\dots H_{i_{n}}
\prod_{1 \le \alpha < \beta \le n}
\left(
1-\frac{\eta^{2}}{(x_{i_{\alpha}}- x_{i_{\beta}})^{2} }
\right)^{-1}
=C_{n}(\{M_a\}),
 \label{eqofeigen}
\end{align}
where $n=1,2, \ldots , L$ and
\beq\label{CCCM}
\begin{array}{lll}
C_{n}(\{M_a\})&=&\displaystyle{
\frac{1}{2\pi i}\oint_{|z|=1}\prod_{a=1}^{K}
(1+zg_a)^{M_a} \, z^{-n-1} dz}
\\ && \\
&=&\displaystyle{\sum_{ n_{1},\dots, n_{K} \in {\mathbb Z}_{\ge 0},
\atop \sum_{j=1}^{K}n_{j}=n}
\binom{M_{1}}{n_{1}}
\ldots
\binom{M_{K}}{n_{K}} \prod_{a=1}^{K}g_{a}^{n_a}}.
\end{array}
\eeq
There are $L$ equations for $L$ unknown quantities $H_1, \ldots , H_L$.
Examples for small values of $L$ are given in the appendix.

Here we point out some simple general properties of these equations.
\begin{enumerate}
\item
This system does not depend on $N|M$ and
the grading parameters
and is also invariant under the following transformations:
a) $\eta \to -\, \eta \,, \,\,$ b)
$\{x_i\}\, \to \, \{-\, x_i\}\,, \,\,$
c) $\, \{H_i\}\to \{-H_i\}$
simultaneously with  $\{g_a\}\to \{-g_a\}$.
\item
For $M_a=L\delta_{aa_1}$ (in this case
$C_n=\frac{L!}{n! (L-n)!}\, g_{a_1}^n$)
there are two distinguished solutions
\beq\label{solsol}
H_j = g_{a_1}\prod_{k=1, \neq j}^L
\left (1\pm \frac{\eta}{x_j \! - \! x_k}\right )
\eeq
which give the eigenvalues of the Hamiltonians on the vector
$(\vv_{a_1})^{\otimes L}$. The two choices of sign
correspond to two possible values of the
grading parameter $\ps (a_1)$.
\item
Assume that $L\leq K$ and
fix $\{ a_1, \ldots , a_L\} \subseteq \{ 1, \ldots , K\}$ such that
all the $a_i$'s are distinct. Set $M_a =\sum_{i=1}^{L}\delta_{aa_i}$,
then the right hand sides of equations (\ref{eqofeigen}) are
elementary symmetric polynomials\footnote{The
elementary symmetric polynomials are defined by means of the
generating function as follows:
$$
\prod_{i=1}^{\mathcal{N}}(1+y_i z)=\sum_{k=0}^{\mathcal{N}}
e_k(y_1, \ldots , y_{\mathcal{N}})z^k.
$$}
$e_k(g_{a_1}, \ldots , g_{a_L})$ of the twist parameters
$g_{a_1}, \ldots , g_{a_L}$.
Then the system of equations (\ref{eqofeigen}) has
$L!$ solutions (counted with multiplicities). Indeed,
at $\eta =0$ the system is just
\begin{align}
e_n(H_1, \ldots , H_L)&= e_n(g_{a_1}, \ldots , g_{a_L}),
\quad n=1, \ldots , L
 \label{eqeta0}
\end{align}
All solutions of this system are given by all possible permutations
of the set $(g_{a_1}, \ldots g_{a_L})$ containing $L$ elements.
\end{enumerate}
The detailed
structure of solutions to (\ref{eqofeigen})
and their correspondence with spectra
of particular spin chains is a subject of further study.
Some examples are discussed in Appendix C.

One can consider the system (\ref{eqofeigen}) with the
right hand sides being ``in general position'' meaning that
there are $L$ twist parameters $g_i$ which are all distinct.
This is the generic situation from which all other possible cases
can be obtained by merging some of the twist parameters.
The generic system (\ref{eqofeigen}) contains $L$ equations
of the form
\begin{align}
\sum_{1 \le i_{1} < \ldots < i_{n} \le L}
 H_{i_{1}}\dots H_{i_{n}}
\prod_{1 \le \alpha < \beta \le n}
\left(
1-\frac{\eta^{2}}{(x_{i_{\alpha}}- x_{i_{\beta}})^{2} }
\right)^{-1}
=e_n(g_{1}, \ldots , g_{L}).
 \label{eqofeigen000}
\end{align}
The complete information about spectra of the
Hamiltonians for $L$-site spin chains
based on all (super)algebras of the type $gl(N|M)$
is contained in the {\it universal spectral variety}
\beq\label{universpectral}
\mbox{{\large $\mathbb{S}_L$}} =\Bigl \{
\bigl (H_1, \ldots H_L; x_1, \ldots , x_L; g_1, \ldots , g_L \bigr )
\Bigm | \mbox{Equations (\ref{eqofeigen000})} \Bigr \}.
\eeq
It is a $2L$-dimensional affine variety embedded into $\mathbb{C}^{3L}$.
The spectra of the Hamiltonians for particular spin chains
are obtained by intersecting with the hyperplanes with fixed
values of $x_i$'s and $g_i$'s.
The variety $\mathbb{S}_L$ is not compact.
We anticipate that a proper compactification
of the universal spectral variety encodes information about the
spectra of Hamiltonians for spin chains when some or all
$x_i$'s coalesce.



\section{The QC correspondence via nested Bethe ansatz}
\setcounter{equation}{0}

In this section we give a direct proof of the QC correspondence based on
the nested Bethe ansatz solution to the susy-XXX spin chains.

\subsection{Bethe ansatz solution for $Y(gl(N|M))$ spin chain}

Here we specify the general nested Bethe ansatz results
(\ref{s06a}), (\ref{s08a}) for the spin chain with vector
representations at the sites. The eigenstates of the T-operator $\Tf
(x)$ are obtained from the reference state $(\mathbf{v}_1)^{\otimes
L}$ with $M_a=L\delta_{a1}$ by action of creation operators. In the
weight space $\mathcal{V}(M_1, \ldots , M_K)$ with $M_1=L-L_1$,
$M_2=L_1-L_2$,  $M_3=L_2-L_3$, $\ldots \,\, $,
$M_{K-1}=L_{K-2}-L_{K-1}$, $M_K=L_{K-1}$ such that $L_1\geq L_2\geq
\ldots \geq L_K$ the eigenvalues of $\mathbf{T}(x)$ are given by the
formula \beq\label{s06}
\begin{array}{lll}
\mathrm{T}(x)&=&\displaystyle{(-1)^{\ps (1)} g_1 \prod_{l=1}^L
\frac{x-x_l +(-1)^{\ps (1)}\eta}{x-x_l}
\prod_{\gamma =1}^{L_1}
\frac{x-\mu_{\gamma}^{1}-(-1)^{\ps (1)}\eta}{x-\mu_{\gamma}^{1}}}
\\ && \\
&& +\,\, \displaystyle{\sum_{b=2}^{K}(-1)^{\ps (b)}g_b
\prod_{\alpha =1}^{L_{b-1}}
\frac{x-\mu_{\alpha}^{b-1} +(-1)^{\ps (b)}\eta}{x-\mu_{\alpha}^{b-1}}
\prod_{\gamma =1}^{L_b}
\frac{x-\mu_{\gamma}^b -(-1)^{\ps (b)}\eta}{x-\mu_{\gamma}^b}},
\end{array}
\eeq
where the parameters $\mu_{\gamma}^b$
(the Bethe roots) obey the system of Bethe
equations
\beq\label{s08}
  \begin{array}{c}
  \displaystyle{
{g_b}\,\prod\limits_{k=1}^L
\frac{\mu^b_\be-x_k+\delta_{b1}
(-1)^{{\mathsf p}(b)}\eta}{\,\,\mu^b_\be-x_k}
 \,\prod\limits_{\ga=1}^{L_{b-1}}
 \frac{\mu^b_\be-\mu_\ga^{b-1}+
 (-1)^{\ps(b)}\eta}{\mu^b_\be-\mu_\ga^{b-1} }
 }
 \\ \ \\
  \displaystyle{\,\, =
 {g_{b+1}}\prod\limits_{\ga\neq \be}^{L_{b}}
\frac{\mu^b_\be-\mu_\ga^{b}+(-1)^{\ps(b+1)}\eta
}{\mu^b_\be-\mu_\ga^{b}-(-1)^{\ps(b)}\eta}
\,\prod\limits_{\ga=1}^{L_{b+1}}\frac{\mu^b_\be-\mu_\ga^{b+1}-
(-1)^{\ps(b+1)}\eta }{\mu^b_\be-\mu_\ga^{b+1}}}\,.
\end{array}
  \eeq
Here $b$ runs from $1$ to $K-1=N+M-1$ and the convention
$L_0=L_K=0$ is implied.
%
Note that the first product in the
r.h.s. of (\ref{s08}) disappears
if $(-1)^{\ps(b)+\ps(b+1)}=-1$.
The Bethe equations are equivalent to the conditions
that $\mathrm{T}(x)$ given by (\ref{s06}) is regular
at the points $x=\mu_\be^b$ for all
$\be =1,...,L_b\,,\ \ b=1,...,K\! -\! 1$.
The corresponding eigenvalues of the Hamiltonians
$\mathbf{H}_i$ are
\beq\label{s10}
 \begin{array}{c}
H_i \left (\{ x_i \}_L\,, \{\mu _{\alpha}^{1}
\}_{L_1},g_1 \right )=
\displaystyle{
 g_1
\prod\limits_{k=1}^L \frac{x_i-x_k+\! (-1)^{\ps (1)}
\eta}{x_i-x_k}
\,\prod\limits_{\ga=1}^{\,L_{1}}\frac{x_i-\mu_\ga^{1} -
(-1)^{\ps(1)}\eta}{x_i-\mu_\ga^{1}}\,, }
\end{array}
  \eeq
 where $\{ x_i \}_L$ emphasizes the dependence on $L$ variables $x_i$
($\{ x_i \}_L$ means $\{ x_i \}_{i=1}^{L}$, in particular,
$\{ x_i \}_0=\emptyset $)
and similarly for $\{\mu _{\alpha}^{1} \}_{L_1}$.

\paragraph{Example: $N+M=2$.}
In the weight space $\mathcal{V}(M_1, M_2)$
with $M_1\geq M_2$
the Bethe equations are:
 \beq\label{s11}
gl(2|0):\ \ \ \
g_1\,\prod\limits_{k=1}^L\frac{\mu^1_\al-x_k+\eta}{\mu^1_\al-x_k}
=g_2 \prod\limits_{\ga\neq\al}^{M_2}
\frac{\mu^1_\al-\mu^1_\ga+\eta}{\mu^1_\al-\mu^1_\ga-\eta}\,,
  \eeq
 \beq\label{s12}
gl(1|1)
\quad  \text{with} \quad (\ps(1),\ps(2))=(0,1)
:\ \ \ \ \ \
g_1\,\prod\limits_{k=1}^L\frac{\mu^1_\al-x_k+\eta}{\mu^1_\al-x_k}
=g_2\,, \qquad \qquad \qquad \ \ \
  \eeq
 \beq\label{s13}
gl(0|2):\ \ \ \
g_1\,\prod\limits_{k=1}^L\frac{\mu^1_\al-x_k-\eta}{\mu^1_\al-x_k}
=g_2 \prod\limits_{\ga\neq\al}^{M_2}
\frac{\mu^1_\al-\mu^1_\ga-\eta}{\mu^1_\al-\mu^1_\ga+\eta}\,.
  \eeq
where $\al=1\,,...\,,M_2$. Note that the equations for
$gl(0|2)$ are obtained from those for $gl(2|0)$ by the
transformation $\eta \to -\eta$.
The spectrum is given by (\ref{s10}) with $\ps (1)=0$ for
$gl(2|0)$, $gl(1|1)$:
 \beq\label{s14}
 \begin{array}{c}
\displaystyle{H_i= g_1\, \prod\limits_{k=1}^L
\frac{x_i-x_k+\eta}{x_i-x_k}
\,\prod\limits_{\ga=1}^{\,M_{2}}\frac{x_i-\mu_\ga^{1} -
\eta}{x_i-\mu_\ga^{1}}\,,
 }
\end{array}
  \eeq
and $\ps(1)=1$ for $gl(0|2)$ which leads to the same expression
with $\eta \to -\eta$.

\subsection{The QC correspondence: a direct proof}

Here we extend the result of \cite{GZZ} to supersymmetric spin
chains.
 \begin{theor}
Substitute
 \beq\label{s21}
 \begin{array}{c}
 \displaystyle{
\dot x_i=-\eta\, H_i\left (\{ x_i \}_L\,, \{\mu _{\alpha}^{1}
\}_{L_1},g_1 \right )\,,\ \ i=1,...,L
 }
 \end{array}
  \eeq
into the Lax matrix for the RS model (\ref{CM6}), i.e.
 consider the matrix
 \beq\label{s211}
 \begin{array}{c}
 \displaystyle{
\left({\mathsf Z}_0\right)_{ij}=\frac{\eta\, H_i}{x_j-x_i+\eta}
 }
 \end{array}
  \eeq
  (see (\ref{QC1})),
where $H_j$ are eigenvalues (\ref{s10}) of the non-local
Hamiltonians of the inhomogeneous
graded
$gl(N|M)$ spin chain on $L$-sites with $N+M=K\leq L$
and the set $\{\mu _{\alpha}^{1} \}_{L_1}$ is taken from any
solution $\{\mu _{\alpha}^{b} \}_{L_b}$, $b=1, \ldots , K-1$
of the Bethe equations (\ref{s08}). Then the spectrum of
the Lax matrix (\ref{s211}) is of the form (\ref{QC2}):
 \beq\label{s22}
 \begin{array}{c}
  \displaystyle{
 \hbox{\rm Spec} \, {\mathsf Z}_0
 \Bigr |_{BE}}
 \displaystyle{=\, \big (\underbrace{g_1\,,\ldots\,,g_1}_{L-L_1}\,, \, \underbrace{g_2\,,\ldots\,,g_2}_{L_1-L_2}\,,\ldots\,,
 \, \underbrace{g_{K\!-\!1}\,,\ldots\,,
 g_{K\!-\!1}}_{L_{K\!-\!2}-L_{K\!-\!1}}\,,\,
 \underbrace{g_K\,,\ldots\,,g_K}_{L_{K\!-\!1}}\big)\,.}
 \end{array}
  \eeq
 \end{theor}

\noindent
\paragraph{Proof.} The proof involves three steps. First, we recall the
proof for the $gl(K|0)$ case. Next, we show how to modify it for the
$gl(0|K)$ case. Lastly, the general case is processed by gluing
together the previous two proofs in a proper way. When odd and even
grading parameters are intermixed then one should apply the scheme
of the proof given below each time the grading is changing.

Instead of $\mathsf{Z}_0$ it is more convenient to deal with
the transposed Lax matrix
$\mathsf{Z}^{\mathsf{t}}_0$ given by
 \beq\label{s20}
 \begin{array}{c}
 \displaystyle{
(\mathsf{Z}^{\mathsf{t}}_0)_{ij}
(\{\dot{x_{k}}\}_L,
 \{x_{k}\}_L, \eta)=-\frac{\dot x_j}{x_i-x_j+\eta}=
 \frac{\eta\, H_j}{x_i-x_j+\eta}\,,\ \ \
 i\,,j=1\,,...\,,L.
 }
 \end{array}
  \eeq
Its spectrum coincides with that of ${\mathsf Z}_0$.

\vskip2mm

\noindent
1. \underline{$gl(K|0)$ case.} The proof given in
\cite{GZZ} is based on the identity
 \beq\label{s23}
\begin{array}{c}
 \det\limits_{L\times L}
 \Bigl ({\mathcal{Z}}\left (\{x_i\}_L,\{y_i \}_{\tilde L},g\right )
 -\la \1 \Bigr  )=(g-\la)^{L-\tilde L}
\det\limits_{\tilde L \times \tilde L}\Bigl ({\widetilde {\mathcal{Z}}}
\left (\{y_i \}_{\tilde L},\{x_i\}_{L},g \right )-\la \1 \Bigr )\,
\end{array}
  \eeq
for the pair of $L\times L$ and $\tilde L\times \tilde L$
matrices
 \beq\label{s24}
 {\mathcal{Z}}_{ij}(\{x_k\}_L,\{y_k \}_{\tilde L},g)=
 \frac{g\,\eta}{x_i-x_j+\eta}\prod\limits_{k\neq j}^L
 \frac{x_j-x_k+\eta}{x_j-x_k}
\prod\limits_{\ga=1}^{\tilde L}\frac{x_j-y_\ga}{x_j-y_\ga+\eta}
  \eeq
and
 \beq\label{s25}
{\widetilde {\mathcal{Z}}}_{\al\be}(\{y_i \}_{\tilde L},\{x_i\}_L,g)=
\frac{g\,\eta}{y_\al-y_\be+\eta}\prod\limits_{\ga\neq
\be}^{\tilde L}\frac{y_\be-y_\ga-\eta}{y_\be-y_\ga}
\prod\limits_{k=1}^L\frac{y_\be-x_k}{y_\be-x_k-\eta}
  \eeq
  (here $L\geq \tilde L$).
In addition, we have
 \beq\label{s26}
 \begin{array}{c}
 \displaystyle{
 \det\limits_{L\times L}
 \Bigl ({\mathcal{Z}^0}(\{x_i\}_L,g)-\lambda \1 \Bigr )=
 \det\limits_{L\times L}\Bigl (
 {{\widetilde {\mathcal{Z}}}^0}(\{y_i\}_L,g)-
 \lambda \1 \Bigr )=(g-\la)^{L}\,,
 }
 \end{array}
  \eeq
where
 \beq\label{s27}
{\mathcal{Z}}_{ij}^0(\{x_k\}_L,g)=
{\mathcal{Z}}_{ij}(\{x_k\}_L,\{y_k \}_0,g)=
 \frac{g\,\eta}{x_i-x_j+\eta}\prod\limits_{k\neq j}^L
 \frac{x_j-x_k+\eta}{x_j-x_k}
  \eeq
and
 \beq\label{s28}
{\widetilde {\mathcal{Z}}}^0_{\al\be}(\{y_i \}_L,g)={\widetilde
{\mathcal{Z}}}_{\al\be}(\{y_i \}_L,\{x_i\}_0,g)=
\frac{g\,\eta}{y_\al-y_\be+\eta}\prod\limits_{\ga\neq
\be}^L\frac{y_\be-y_\ga-\eta}{y_\be-y_\ga}\,.
  \eeq
The idea is to calculate $\det(\mathsf{Z}_0-\la \1)$ by
sequential usage of the identity (\ref{s23}) (which allows one
to pass to a
smaller matrix) and the Bethe equations (BE) (\ref{s08})
with $\ps(b)=0$ for
all $b$.
Schematically\footnote{Here we symbolically write
simply $\bigl <{\mathcal Z}\bigr >$ for
the characteristic polynomial
$\det ({\mathcal Z}-\lambda \1)$
with some overall factor.},
the procedure of the proof is as follows:
 \beq\label{s29}
 \begin{array}{c}
 \displaystyle{
\bigl < \mathsf{Z}^{\mathsf{t}}_0\bigr >
\left (-{\eta}\left  \{ H_j\right \}_L,
 \left \{ x_j\right \} _L, \, \eta \right )\stackrel{(\ref{s10})}{=}
 \bigl <{\mathcal{Z}}\bigr >\,(\{x_i-\eta\}_L,\{\mu_\al^1
 \}_{L_1},g_1)\stackrel{(\ref{s23})}{\to}
 }
 \\ \ \\
 \displaystyle{
\bigl <{\widetilde {\mathcal{Z}}}\bigr >\,(\{\mu_\al^1
 \}_{L_1},\{x_i-\eta\}_L,g_1)
 \stackrel{\hbox{BE}_{b=1}}{=}\bigl <{\mathcal{Z}}\bigr >\,
 (\{\mu_{\alpha}^{1}-\eta\}_{L_1},\{
\mu_{\alpha}^{2}\}_{L_2},g_2)\stackrel{(\ref{s23})}{\to}
 }
  \\ \ \\
 \displaystyle{
\bigl <{\widetilde {\mathcal{Z}}}\bigr >\,(\{
\mu_{\alpha}^{2}\}_{L_2},\{\mu_{\alpha}^{1}-\eta\}_{L_1},g_2)
\stackrel{\hbox{BE}_{b=2}}{=}\bigl <{\mathcal{Z}}\bigr >\,
(\{\mu_{\alpha}^{2}-\eta\}_{L_2},\{
\mu_{\alpha}^{3}\}_{L_3},g_3)\stackrel{(\ref{s23})}{\to}...
 }
 \end{array}
  \eeq
Each time we use (\ref{s23}) the characteristic polynomial
$\det(\mathcal{Z} -\la \1)$ acquires the factor
$(g_b-\la )^{L_{b-1}-L_b}$ except for the last step when we use
(\ref{s26}) to get $(g_K -\la)^{L_{K-1}}$.

\vskip2mm

\noindent
2. \underline{$gl(0|K)$ case.} It is easy to see that this
case is similar to the previous one but the roles of ${\mathcal{Z}}$
and ${\widetilde {\mathcal{Z}}}$ in the scheme (\ref{s29}) get
interchanged:
 \beq\label{s30}
 \begin{array}{c}
 \displaystyle{
\bigl <\mathsf{Z}^{\mathsf{t}}_0
\bigr >\left (-{\eta}\left  \{ H_j\right \}_L,
 \left \{ x_j\right \} _L, \, \eta \right )
 \stackrel{(\ref{s10})}{=}\bigl <{\widetilde {\mathcal{Z}}}\bigr >\,
 (\{x_i+\eta\}_L,\{\mu_\al^1
 \}_{L_1},g_1)\stackrel{(\ref{s23})}{\to}
 }
 \\ \ \\
 \displaystyle{
\bigl <{{\mathcal{Z}}}\bigr >\,(\{\mu_\al^1
 \}_{L_1},\{x_i+\eta\}_L,g_1)\stackrel{\hbox{BE}_{b=1}}{=}
\bigl <{\widetilde {\mathcal{Z}}}\bigr >
\,(\{\mu_{\alpha}^{1}+\eta\}_{L_1},\{
\mu_{\alpha}^{2}\}_{L_2},g_2)\stackrel{(\ref{s23})}{\to}
 }
  \\ \ \\
 \displaystyle{
\bigl <{ {\mathcal{Z}}}\bigr >\,(\{
\mu_{\alpha}^{2}\}_{L_2},\{\mu_{\alpha}^{1}+\eta\}_{L_1},g_2)\stackrel{\hbox{BE}_{b=2}}{=}
\bigl <{\widetilde{\mathcal{Z}}}\bigr >
\,(\{\mu_{\alpha}^{2}+\eta\}_{L_2},\{
\mu_{\alpha}^{3}\}_{L_3},g_3)\stackrel{(\ref{s23})}{\to}...
 }
 \end{array}
  \eeq
Here we use the BE (\ref{s08})
with $\ps(b)=1$ for
all $b$.

\vskip2mm

\noindent
3. \underline{$gl(N|M)$ case.} We assume that $N,M\geq 1$. The
scheme (\ref{s29}) works for $b=1,...,N-1$ while (\ref{s30}) does for
$b=N+1,...,N+M$. In order to switch from the scheme (\ref{s29}) to
(\ref{s30}) we
need an intermediate step. It is accomplished by the BE
(\ref{s08}) at $b=N$:
 \beq\label{s31}
  \begin{array}{c}
  \displaystyle{
\hbox{BE}_{b=N}:\ \ \
{g_N}\,\prod\limits_{\ga=1}^{L_{N-1}}\frac{\mu^N_\be-\mu_\ga^{N-1}+
 \eta}{\mu^N_\be-\mu_\ga^{N-1}} =
 {g_{N+1}}
\,\prod\limits_{\ga=1}^{L_{N+1}}\frac{\mu^N_\be-\mu_\ga^{N+1}+\eta
}{\mu^N_\be-\mu_\ga^{N+1}}\,.
 }
\end{array}
  \eeq
Then for
 \beq\label{s32}
  \begin{array}{c}
  \displaystyle{
\bigl <{\widetilde {\mathcal{Z}}}\bigr >\,(\{
\mu_{\alpha}^{N}\}_{L_N},\{\mu_{\alpha}^{N-1}-\eta\}_{L_{N-1}},g_N)
 }
 \\ \ \\
  \displaystyle{=
\frac{g_N\,\eta}{\mu_\al^N-\mu_\be^N+\eta}
 \prod\limits_{\ga\neq \be}^{L_N}\frac{\mu_\be^N-\mu_\ga^N-\eta}{\mu_\be^N-\mu_\ga^N}
 \prod\limits_{\ga=1}^{L_{N-1}}\frac{\mu_\be^N-\mu_\ga^{N-1}+\eta}{\mu_\be^N-\mu_\ga^{N-1}}
 }
\end{array}
  \eeq
we have:
 \beq\label{s33}
  \begin{array}{c}
  \displaystyle{
\bigl <{\widetilde {\mathcal{Z}}}\bigr >\,(\{
\mu_{\alpha}^{N}\}_{L_N},\{\mu_{\alpha}^{N-1}-\eta\}_{L_{N-1}},g_N)\stackrel{(\ref{s31})}{=}
\bigl <{\widetilde {\mathcal{Z}}}\bigr >\,(\{
\mu_{\alpha}^{N}+\eta\}_{L_N},\{\mu_{\alpha}^{N+1}\}_{L_{N+1}},g_{N+1})\,.
 }
\end{array}
  \eeq
This finishes the proof. $\blacksquare$

\section{Concluding remarks}

Lastly, we would like to point out some unsolved problems
and directions for further research.
\begin{itemize}
\item[a)]
The properly taken limit
$\eta \to 0$ should lead to the similar QC correspondence between graded
quantum Gaudin models and classical Calogero-Moser systems.
This is technically involved but rather straightforward procedure.
The details will be published elsewhere.
\item[b)] A less straightforward but quite realistic program
is the extension to the models based on trigonometric solutions
to the (graded) Yang-Baxter equation (the graded XXZ magnets
and corresponding vertex models of statistical mechanics).
As the results of \cite{Z12} suggest, one can
expect the trigonometric RS model on the classical side
of the QC correspondence. The precise relation between
eigenvalues of the Lax matrix and the twist parameters
of the spin chain is to be
elaborated.
\item[c)] The extension to quantum
integrable models with elliptic $R$-matrices
is problematic. Conceivably
this might require new ideas.
At the same time, the most natural candidate for
the classical part of the QC correspondence is the elliptic RS model.
The role of the spectral parameter which enters its Lax matrix
is to be clarified.
\item[d)] There is a well-known duality \cite{Ruijsdual,FGNR}
of the classical
RS (and Calogero-Moser) type models when
the action variables in a given
system are treated as coordinates of particles in the dual one.
Equivalently, the
soliton-like tau-function whose zeros move as the RS particles
becomes the spectral determinant for the dual system and vice versa.
An interesting future perspective is to realize the meaning
of this duality in the context of the quantun spin chains.
Presumably, this duality implies some correspondence
between spectra of different spin chains.
\end{itemize}

\section*{Appendix A: The higher T-operators through
supercharacters}
\label{higherT}
\addcontentsline{toc}{section}{Appendix A:
The higher T-operators through
supercharacters}
\def\theequation{A\arabic{equation}}
\def\theHequation{\theequation}
\setcounter{equation}{0}

Here we show how to derive (\ref{ntl3}).
We have:
{\small
$$
\Tf _{\lambda}(x)=\mathrm{str}_{V_\lambda}\left [
\Bigl (\mathbf{I}+\frac{\eta}{x\! -\! x_L}\,
\mathbf{P}_{\lambda}^{0L}\Bigr )\ldots
\Bigl (\mathbf{I}+\frac{\eta}{x\! -\! x_2}\,
\mathbf{P}_{\lambda}^{02}\Bigr )
\Bigl (\mathbf{I}+\frac{\eta}{x\! -\! x_1}\,
\mathbf{P}_{\lambda}^{01}\Bigr )
\bigl ( \pi_{\lambda}(\g )\otimes \mathbf{I}\bigr )
\right ]
$$
$$
=\, \mathrm{str}_{V_\lambda}\pi_{\lambda}(\g ) \, \mathbf{I}
+\,
\sum_{j} \frac{\eta}{x\! -\! x_j}\,
\mathrm{str}_{V_\lambda}
\Bigl (\mathbf{P}_{\lambda}^{0j}
( \pi_{\lambda}(\g )\otimes \mathbf{I}\bigr )
\Bigr )\,
 +\,
\sum_{i<j}\frac{\eta^2}{(x\! -\! x_i)(x\! -\! x_j)}\,
\mathrm{str}_{V_\lambda}
\Bigl (\mathbf{P}_{\lambda}^{0j}
\mathbf{P}_{\lambda}^{0i}
( \pi_{\lambda}(\g )\otimes \mathbf{I}\bigr )
\Bigr )
$$
$$
+ \,\,\,\, {\bf \ldots}\,\, \, + \,
\frac{\eta^L}{(x\! -\! x_1)\ldots (x\! -\! x_L)}\,
\mathrm{str}_{V_\lambda}
\Bigl (\mathbf{P}_{\lambda}^{0L}
\ldots  \mathbf{P}_{\lambda}^{01}
( \pi_{\lambda}(\g )\otimes \mathbf{I}\bigr )
\Bigr ).
$$
}
Plugging here the explicit form of $\mathbf{P}_{\lambda}^{0j}$
(\ref{Plam}), we get:
{\small
$$
\Tf _{\lambda}(x)=\mathrm{str}_{V_\lambda}\pi_{\lambda}(\g )
\, \mathbf{I} +\,
\sum_{j} \sum_{ab} \frac{\eta (-1)^{\ps (b)}
\e_{ba}^{(j)}}{x\! -\! x_j}\,
\mathrm{str}_{V_\lambda}
\Bigl ( \pi_{\lambda}(\e_{ab})\pi_{\lambda}(\g )\Bigr )
$$
$$
+ \, \sum_{i<j} \sum_{a_1, b_1; a_2, b_2}
\frac{\eta^2 (-1)^{\ps (b_1)+\ps (b_2)}
\e_{b_1 a_1}^{(i)}\e_{b_2 a_2}^{(j)}}{(x\! -\! x_1)(x\! -\! x_2)}\,
\mathrm{str}_{V_\lambda}
\Bigl ( \pi_{\lambda}(\e_{a_2b_2})
\pi_{\lambda}(\e_{a_1b_1})\pi_{\lambda}(\g )\Bigr )
$$
$$
\begin{array}{l}
\,\,\,\,\,\,\,\, + \, {\bf \ldots}\,
\\ \\
\displaystyle{+ \,\,
\sum_{\{a_1, b_1; \ldots ; a_L,b_L\}}
\frac{\eta^L (-1)^{\ps (b_1)+\ldots +\ps (b_L)}
\e_{b_1a_1}^{(1)} \ldots \e_{b_La_L}^{(L)}}{(x\! -\! x_1)
\ldots (x\! -\! x_L)}\,
\mathrm{str}_{V_\lambda}
\Bigl ( \pi_{\lambda}(\e_{a_Lb_L})\ldots
\pi_{\lambda}(\e_{a_1b_1})\pi_{\lambda}(\g )\Bigr )}\,.
\end{array}
$$
}

The last step is based
on the following simple lemma:

\noindent
{\bf Lemma A1.}
{\it Let $\pi$ be a representation of
$U(gl(N|M))$, $\chi$ its character and
$\mathbf{h}_1, \mathbf{h}_2, \ldots ,
\mathbf{h}_n$ be any homogeneous elements of
the superalgebra $gl(N|M)$ such that
$[\mathbf{h}_i, \mathbf{h}_i]=0$ (here $[\, , \, ]$ means
the graded commutator). Then for any group element
$\g\in GL(N|M)$ it holds:
\beq\label{A1}
\mathrm{str}
\Bigl [ \pi (\mathbf{h}_1)\pi (\mathbf{h}_2)\ldots
\pi (\mathbf{h}_n)\pi (\g )\Bigr ]=
\frac{\p}{\p \varepsilon_n}
\ldots \frac{\p}{\p \varepsilon_1}
\chi \bigl ( e^{\varepsilon_1\mathbf{h}_1}\ldots
e^{\varepsilon_n \mathbf{h}_n}\g \bigr )
\Biggm |_{\varepsilon_i=0},
\eeq
where it is implied that $\ps (\varepsilon_i)=\ps (\mathbf{h}_i)$.
}

\noindent
The proof is simple. Since $[\mathbf{h}_i, \mathbf{h}_i]=0$,
the exponents $e^{\varepsilon_i\mathbf{h}_i}$ are supergroup
elements for any $\varepsilon_i$.
Therefore, we have the chain of equalities
$$
\pi \bigl (e^{\varepsilon_1\mathbf{h}_1}\ldots
e^{\varepsilon_n\mathbf{h}_n}\g \bigr )=
\Bigl [\overrightarrow{\prod_{j=1}^{n}}\pi \bigl (
e^{\varepsilon_j\mathbf{h}_j}\bigr )\Bigr ] \pi (\g )=
\Bigl [\overrightarrow{\prod_{j=1}^{n}}
e^{\varepsilon_j \pi (\mathbf{h}_j)}
\Bigr ] \pi (\g ),
$$
from which it follows directly that
$$
\pi (\mathbf{h}_1)\pi (\mathbf{h}_2)\ldots
\pi (\mathbf{h}_n)\pi (\g )=
\frac{\p}{\p \varepsilon_n}
\ldots \frac{\p}{\p \varepsilon_1}
\pi \bigl ( e^{\varepsilon_1\mathbf{h}_1}\ldots
e^{\varepsilon_n \mathbf{h}_n}\g \bigr )
\Biggm |_{\varepsilon_i=0}.
$$
Taking supertrace of the both sides, we obtain (\ref{A1}).
Applying the lemma to the case
$\mathbf{h}_i =\e_{a_ib_i}$, we arrive at
(\ref{ntl3}).

\section*{Appendix B: Hamiltonian formulation of the RS model}
\label{Ham}
\addcontentsline{toc}{section}{Appendix B:
Hamiltonian formulation of the RS model}
\def\theequation{B\arabic{equation}}
\def\theHequation{\theequation}
\setcounter{equation}{0}

For completeness, we give here the Hamiltonian formulation of the
$L$-particle RS model, including the higher flows.
The Hamiltonian is
\beq\label{RS1}
{\cal H}_1=\sum_{i=1}^L e^{-\eta p_i}\prod_{k=1, \neq i}^L
\frac{x_i-x_k+\eta}{x_i-x_k}\,,
\eeq
with $\{p_i, x_i\}$ being the canonical variables with the
standard Poisson brackets.
The Hamiltonian equations of motion
$\displaystyle{\left ( \begin{array}{c}\dot x_i \\
\dot p_i \end{array} \right )
=\left ( \begin{array}{cc} \p_{p_i}{\cal H}_1\\
-\p_{x_i}{\cal H}_1\end{array} \right )}$ give the
connection between velocity and momentum
\beq\label{RS2a}
\dot x_i=-\eta e^{-\eta p_i}\prod_{k=1, \neq i}^L
\frac{x_i-x_k+\eta}{x_i-x_k}
\eeq
and the equations of motion (\ref{CM9}).

The RS model is known to be integrable, with the higher integrals
of motion in involution being given by ${\cal H}_k=
\mbox{tr}\, {\sf Z}^k$, where ${\sf Z}$
is the Lax matrix of the model (\ref{CM6}):
\beq\label{RS3}
{\sf Z}_{ij}=\frac{\dot x_i}{x_i\! -\! x_j \! -\! \eta}=
\frac{\eta \, e^{-\eta p_i}}{x_j\! -\! x_i \! +\! \eta}\,\,
\prod_{k=1, \neq i}^L \! \left ( 1+\frac{\eta}{x_i \! -\! x_k}\right ).
\eeq
These integrals of motion can be regarded as Hamiltonians
generating flows in the ``higher times'' $t_k$ via the
Hamiltonian equations
\beq\label{CM12}
\left (\begin{array}{l}\p_{t_k}x_i \\ \p_{t_k} p_i \end{array} \right )=
\left (\begin{array}{r} \p_{p_i} {\cal H}_k \\
- \p_{i} {\cal H}_k \end{array} \right ), \quad k\geq 1.
\eeq
Moreover,
the dynamics in the higher time $t_k$ is precisely
the one induced by the mKP flow on the roots of the
tau function (\ref{tau1}).
The fact that the integrals of motion ${\cal H}_k$
are in involution agrees with the commutativity
of the mKP flows.
The proof is based on the linear problems (\ref{CM5}), (\ref{CM5a})
and general relation (\ref{dif3}).
We will not repeat it here since it
is technically involved. It can be found in
\cite{Iliev,Z13}.

\section*{Appendix C: Examples for small
values of $L$
and limiting cases}
\label{proof-wick}
\addcontentsline{toc}{section}{Appendix C:
Examples for small values of $L$
and limiting cases}
\def\theequation{C\arabic{equation}}
\def\theHequation{\theequation}
\setcounter{equation}{0}

\underline{$L=1$}.
For $L=1$,
$\displaystyle{\Tf_{\Box} (x) =
 (\mathrm{str} \, \g ) \1
+\frac{\eta }{x-x_{1}}\sum_{a=1}^{K}g_{a} \e_{aa}}$
and ${\mathbf H}_{1} =
\sum_{a=1}^{K}g_{a} \e_{aa}=\g$. This case is trivial
because there is only one Hamiltonian which is already diagonal.
The system (\ref{eqofeigen}) is in a trivial agreement with this:
it states that $H_1=g_a$.
The master T-operator has the form
\beq\label{explT1}
e^{-\mathrm{str} \, \xi  \left (\mathbf{t}, \g \right )}\,
\frac{\Ts (x, \tf )}{\Ts _{\emptyset}(x)} \Biggm |_{L=1}=\,
\mathbf{I}+\eta
\sum_{k \ge 1}
\frac{kt_{k}(\g^{(1)})^{k}}{x-x_{i}}.
\eeq

\bigskip

\noindent
\underline{$L=2$}.
The case $L=2$ is more interesting.
We have: $\g = g_1 \e_{11}+g_2 \e_{22}$,
\beq
\begin{array}{lll}
\Tf_{\Box} (x)&=&\displaystyle{(\mathrm{str} \, \g ) \1 \otimes \1
+
\frac{\eta }{x-x_{1}}
\, \g \otimes \1
+
\frac{\eta }{x-x_{2} }}
\\ && \\
&& \displaystyle{\phantom{aaa} +\,
\, \frac{\eta^{2}}{(x-x_{1})(x-x_{2}) }
\sum_{a,b=1}^{K} (-1)^{{\mathsf p} (a)}
g_{b} \, \e_{ba} \otimes \e_{ab}\,. }
\end{array}
\eeq
It is convenient to work in the basis $\vv_{a}\otimes \vv_{b}$
of the space $\mathcal{V}=\mathbb{C}^{N|M}\otimes \mathbb{C}^{N|M}$.
The Hamiltonians act as follows:
\begin{align}
{\mathbf H}_{1} \vv_{a} \otimes \vv_{b} &=
 g_{a}   \vv_{a} \otimes \vv_{b}
+
\frac{(-1)^{ \ps(a) \ps(b)}\eta g_{b}}{x_{1}-x_{2} }
 \vv_{b} \otimes \vv_{a},
\\
{\mathbf H}_{2} \vv_{a} \otimes \vv_{b} &=
 g_{b}   \vv_{a} \otimes \vv_{b}
+
\frac{(-1)^{ \ps(a) \ps(b)}\eta  g_{b}}{x_{2}-x_{1} }
 \vv_{b} \otimes \vv_{a},
\end{align}
where we have used the rule
$$(\e_{cd} \otimes \e_{dc} )(\vv_{a} \otimes \vv_{b})
=(-1)^{\ps(\e_{dc})\ps(\vv_{a} )}\e_{cd} \vv_{a} \otimes \e_{dc}\vv_{b}
=(-1)^{(\ps(d)+\ps(c))\ps(a)}
\delta_{da} \delta_{cb}
\vv_{c} \otimes \vv_{d}.$$
Therefore, eigenvalues of ${\mathbf H}_{1}$ in the subspace
of  $\mathcal{V}$
spanned by the vectors $\vv_{a_1} \otimes \vv_{a_2}$ and
$\vv_{a_2} \otimes \vv_{a_1}$ with some fixed $a_1 \ne a_2$ are given by
diagonalizing the $2 \times 2$ matrix
\[
\begin{pmatrix}
g_{a_1} &
\displaystyle{\frac{(-1)^{ \ps(a_1) \ps(a_2)}
\eta g_{a_1}}{x_{1}-x_{2}}} \\
\displaystyle{\frac{(-1)^{ \ps(a_1) \ps(a_2)}
\eta  g_{a_2}}{x_{1}-x_{2}}} & g_{a_2}
\end{pmatrix} .
\]
Then the spectrum of the operators
 $({\mathbf H}_{1}, {\mathbf H}_{2})$ is
\begin{align}\label{HHH1}
(H_{1},H_{2})&=
\left( g_{a_1} +
\frac{(-1)^{ \ps(a_1)} \eta  g_{a_1}}{x_{1}-x_{2} },\;
g_{a_1} +
\frac{(-1)^{ \ps(a_1)} \eta g_{a_1}}{x_{2}-x_{1} }
\right)
\nonumber
\\ &
 \quad \text{in} \quad {\mathbb C} \vv_{a_1} \otimes \vv_{a_1}
 \quad (M_a=2\delta_{aa_1}),
\\
(H_{1},H_{2})& =
\left( \frac{g_{a_1}\! +\! g_{a_2}\! +\! \sqrt{R}}{2},\;
\frac{g_{a_1}\! +\! g_{a_2}\!  -\!  \sqrt{R}}{2}
\right)
,
\left( \frac{g_{a_1}\! +\! g_{a_2}\! -\! \sqrt{R}}{2},\;
\frac{g_{a_1}\! +\! g_{a_2}\!  +\! \sqrt{R}}{2}
\right)
\nonumber
\\ &
 \qquad \text{in} \quad
{\mathbb C}\vv_{a_1} \otimes \vv_{a_2}+
{\mathbb C}\vv_{a_2} \otimes \vv_{a_1}
\quad \text{for} \quad a_1 \ne  a_2\quad (M_a=\delta_{aa_1}+\delta_{aa_2}).
\label{HHH2}
\end{align}
Here $\displaystyle{R=(g_{a_1}-g_{a_2})^{2}+
\frac{ 4 \eta^{2} g_{a_1}g_{a_2}}{(x_{1}-x_{2})^{2}}}\,\,$ and
$1\leq a_1, a_2\leq K$ are some fixed indices.

Let us compare these results with solutions of the system
(\ref{eqofeigen}) which in our case reduces to
$$
\left \{ \begin{array}{l} H_1+H_2=2g_{a_1} \\ \displaystyle{
H_1H_2=g_{a_1}^2 \Bigl (1-\frac{\eta^2}{(x_1\! -\! x_2)^2}\Bigr )}
\end{array}
\right. \quad \mbox{for $M_a=2\delta_{aa_1}$}
$$
and
$$
\;\;\; \qquad \left \{ \begin{array}{l} H_1+H_2=g_{a_1}
+g_{a_2}\\ \displaystyle{
H_1H_2=g_{a_1} g_{a_2}
\Bigl (1-\frac{\eta^2}{(x_1\! -\! x_2)^2}\Bigr )}
\end{array}
\right. \quad \mbox{for $M_a=\delta_{aa_1}+\delta_{aa_2}$.}
$$
Each system has two solutions for the pair $(H_1, H_2)$.
One can easily check that the two solutions of the former system
are just (\ref{HHH1}) for the two possible values
$\ps (a_1)=0,1$ while the two solutions of the latter one
are given by (\ref{HHH2}).

The master T-operator has the form
\begin{align}
\label{explT2}
&
e^{-\mathrm{str} \, \xi  \left (\mathbf{t}, \g \right )}\,
\frac{\Ts (x, \tf )}{\Ts _{\emptyset}(x)} \Biggm |_{L=2}=\,\,
\mathbf{I}+\eta \sum_{i=1}^{2}
\sum_{k \ge 1}
\frac{kt_{k}(\g^{(i)})^{k}}{x-x_{i}}
\, +\,
\frac{\eta^{2}}{(x-x_{1})(x-x_{2})}
\nonumber
\\
&
\times
\left[
\left(
\sum_{k_{1} \ge 1}
k_{1}t_{k_{1}}
(\g^{(1)})^{k_{1}}
\right)
\left(
\sum_{k_{2} \ge 1}
k_{2}t_{k_{2}}
(\g^{(2)})^{k_{2}}
\right)
+
\sum_{k \ge 1}
kt_{k}
\sum_{\alpha =0}^{k-1}
 {\mathbf P}_{12}
(\g^{(1)})^{\alpha} (\g^{(2)})^{k-\alpha}
\right].
\end{align}

\bigskip

\noindent
\underline{$L=3$}.
We start with the explicit form of the Hamiltonians $\mathbf{H}_i$.
Let us introduce the short-hand notation
$\ps_{i}:=(-1)^{\ps(a_{i})}$, $\ps_{ij}:=(-1)^{\ps(a_{i})\ps(a_{j})}$.
In the 3-dimensional subspace spanned by the vectors
$\vv_{a_{1}} \otimes \vv_{a_{1}} \otimes \vv_{a_{2}}$,
$\vv_{a_{1}} \otimes \vv_{a_{2}} \otimes \vv_{a_{1}}$
and
$\vv_{a_{2}} \otimes \vv_{a_{1}} \otimes \vv_{a_{1}}$
for $a_{1} \ne a_{2}$, we have
\begin{align}
{\mathbf H}_{1}&=
\left(
\begin{array}{ccc}
 g_{a_{1}} \left(\frac{\eta  \ps_{1}}{x_{12}}+1\right) &
\frac{\eta ^2 g_{a_{1}} \ps_{1}\ps_{12}}{x_{12} x_{13}} &
\frac{\eta  g_{a_{1}} \ps_{1}}{x_{13}} \\
 0 & g_{a_{1}} \left(\frac{\eta  \ps_{1}}{x_{13}}+1\right) &
\frac{\eta  g_{a_{1}} \ps_{12} \left(x_{13}+\eta
\ps_{1}\right)}{x_{12} x_{13}} \\
 \frac{\eta  g_{a_{2}} \left(x_{12} \ps_{1}+\eta \right)}{x_{12} x_{13}} &
\frac{\eta  g_{a_{2}} \ps_{12}}{x_{12}} & g_{a_{2}}
\end{array}
\right),
\\[10pt]
{\mathbf H}_{2}&=
\left(
\begin{array}{ccc}
 g_{a_{1}} \left(-\frac{\eta  \ps_{1}}{x_{12}}+1\right) &
\frac{\eta  g_{a_{1}} \ps_{12} \left(x_{12}-\eta
\ps_{1}\right)}{x_{12} x_{23}} & 0 \\
 \frac{\eta  g_{a_{2}} \ps_{12}}{x_{23}} & g_{a_{2}} &
-\frac{\eta  g_{a_{1}} \ps_{12} \left(x_{23}+\eta
\ps_{1}\right)}{x_{12} x_{23}} \\
 \frac{-\eta ^2 g_{a_{2}}}{x_{12} x_{23}} &
-\frac{\eta  g_{a_{2}} \ps_{12}}{x_{12}} &
g_{a_{1}} \left(\frac{\eta  \ps_{1}}{x_{23}}+1\right)
\end{array}
\right),
\\[10pt]
{\mathbf H}_{3}&=
\left(
\begin{array}{ccc}
 g_{a_{2}} & -\frac{\eta  g_{a_{1}} \ps_{12}
 \left(x_{13}-\eta  \ps_{1}\right)}{x_{23} x_{13}} &
-\frac{\eta  g_{a_{1}} \ps_{1}}{x_{13}} \\
 -\frac{\eta  g_{a_{2}}  \ps_{12}}{x_{23}} &
g_{a_{1}} \left(-\frac{\eta  \ps_{1}}{x_{13}}+1\right) &
\frac{\eta ^2 g_{a_{1}} \ps_{1}\ps_{12}}{x_{13} x_{23}} \\
 -\frac{\eta  g_{a_{2}} \left(x_{23} \ps_{1}-\eta \right)}{x_{13} x_{23}} & 0
& g_{a_{1}} \left(-\frac{\eta \ps_{1}}{x_{23}}+1\right)
\end{array}
\right).
\end{align}
In the 6-dimensional subspace spanned by the vectors
$\{\vv_{a_{\tau(1)}} \otimes \vv_{a_{\tau(2)}}
\otimes \vv_{a_{\tau(3)}}\}_{\tau \in S_{3}}$,
where $S_{3}$ is the permutation group over $\{1,2,3\}$ and
 $a_{i} \ne a_{j}$ if $i \ne j$,  we have
\begin{align}
{\mathbf H}_{1}&=
\left(
\begin{array}{cccccc}
 g_{a_{1}} & 0 & \frac{\eta  g_{a_{1}} \ps_{12}}{x_{12}} &
\frac{\eta ^2 g_{a_{1}} \ps_{12}\ps_{13}}{x_{12} x_{13}} & 0 &
\frac{\eta  g_{a_{1}} \ps_{12}\ps_{13}\ps_{23}}{x_{13}} \\
 0 & g_{a_{1}} & 0 &
\frac{\eta  g_{a_{1}} \ps_{12}\ps_{13}\ps_{23}}{x_{13}} &
\frac{\eta  g_{a_{1}} \ps_{13}}{x_{12}} &
\frac{\eta ^2 g_{a_{1}} \ps_{12}\ps_{13}}{x_{12} x_{13}} \\
 \frac{\eta  g_{a_{2}} \ps_{12}}{x_{12}} &
\frac{\eta ^2 g_{a_{2}} \ps_{12}\ps_{23}}{x_{12} x_{13}} &
g_{a_{2}} & 0 &
\frac{\eta  g_{a_{2}} \ps_{12}\ps_{13}\ps_{23}}{x_{13}} & 0 \\
 0 & \frac{\eta  g_{a_{2}} \ps_{12}\ps_{13}\ps_{23}}{x_{13}} &
0 & g_{a_{2}} & \frac{\eta ^2 g_{a_{2}} \ps_{12}\ps_{23}}{x_{12} x_{13}} &
\frac{\eta  g_{a_{2}} \ps_{23}}{x_{12}} \\
 \frac{\eta ^2 g_{a_{3}} \ps_{13}\ps_{23}}{x_{12} x_{13}} &
\frac{\eta  g_{a_{3}} \ps_{13}}{x_{12}} &
\frac{\eta  g_{a_{3}} \ps_{12}\ps_{13}\ps_{23}}{x_{13}} & 0 &
g_{a_{3}} & 0 \\
 \frac{\eta  g_{a_{3}} \ps_{12}\ps_{13}\ps_{23}}{x_{13}} & 0 &
\frac{\eta ^2 g_{a_{3}} \ps_{13}\ps_{23}}{x_{12} x_{13}} &
\frac{\eta  g_{a_{3}} \ps_{23}}{x_{12}} & 0 & g_{a_{3}}
\end{array}
\right),
\end{align}
\begin{align}
{\mathbf H}_{2}&=
\left(
\begin{array}{cccccc}
 g_{a_{2}} & \frac{\eta  g_{a_{2}} \ps_{23}}{x_{23}} & \frac{\eta  g_{a_{1}} \ps_{12}}{x_{21}} &
-\frac{\eta ^2 g_{a_{1}} \ps_{12}\ps_{13}}{x_{12} x_{23}} & 0 & 0 \\
 \frac{\eta  g_{a_{3}} \ps_{23}}{x_{23}} & g_{a_{3}} & 0 & 0 & \frac{\eta  g_{a_{1}} \ps_{13}}{x_{21}} &
-\frac{\eta ^2 g_{a_{1}} \ps_{12}\ps_{13}}{x_{12} x_{23}} \\
 \frac{\eta  g_{a_{2}} \ps_{12}}{x_{21}} & -\frac{\eta ^2 g_{a_{2}} \ps_{12}\ps_{23}}{x_{12} x_{23}} & g_{a_{1}} &
\frac{\eta  g_{a_{1}} \ps_{13}}{x_{23}} & 0 & 0 \\
 0 & 0 & \frac{\eta  g_{a_{3}} \ps_{13}}{x_{23}} & g_{a_{3}} &
-\frac{\eta ^2 g_{a_{2}} \ps_{12}\ps_{23}}{x_{12} x_{23}} &
\frac{\eta  g_{a_{2}} \ps_{23}}{x_{21}} \\
 -\frac{\eta ^2 g_{a_{3}} \ps_{13}\ps_{23}}{x_{12} x_{23}} & \frac{\eta  g_{a_{3}} \ps_{13}}{x_{21}} &
0 & 0 & g_{a_{1}} & \frac{\eta  g_{a_{1}} \ps_{12}}{x_{23}} \\
 0 & 0 & -\frac{\eta ^2 g_{a_{3}} \ps_{13}\ps_{23}}{x_{12} x_{23}}
 & \frac{\eta  g_{a_{3}} \ps_{23}}{x_{21}} &
\frac{\eta  g_{a_{2}} \ps_{12}}{x_{23}} & g_{a_{2}}
\end{array}
\right),
\end{align}
\begin{align}
{\mathbf H}_{3}&=
\left(
\begin{array}{cccccc}
 g_{a_{3}} & \frac{\eta  g_{a_{2}} \ps_{23}}{x_{32}} & 0 &
 -\frac{\eta ^2 g_{a_{1}} \ps_{12}\ps_{13}}{x_{13} x_{32}} & 0 &
\frac{\eta  g_{a_{1}} \ps_{12}\ps_{13}\ps_{23}}{x_{31}} \\
 \frac{\eta  g_{a_{3}} \ps_{23}}{x_{32}} & g_{a_{2}} & 0 &
\frac{\eta  g_{a_{1}} \ps_{12}\ps_{13}\ps_{23}}{x_{31}} & 0 &
-\frac{\eta ^2 g_{a_{1}} \ps_{12}\ps_{13}}{x_{13} x_{32}} \\
 0 & -\frac{\eta ^2 g_{a_{2}} \ps_{12}\ps_{23}}{x_{13} x_{32}} & g_{a_{3}} &
\frac{\eta  g_{a_{1}} \ps_{13}}{x_{32}} &
\frac{\eta  g_{a_{2}} \ps_{12}\ps_{13}\ps_{23}}{x_{31}} & 0 \\
 0 & \frac{\eta  g_{a_{2}} \ps_{12}\ps_{13}\ps_{23}}{x_{31}} &
\frac{\eta  g_{a_{3}} \ps_{13}}{x_{32}} & g_{a_{1}} &
-\frac{\eta ^2 g_{a_{2}} \ps_{12}\ps_{23}}{x_{13} x_{32}} & 0 \\
 -\frac{\eta ^2 g_{a_{3}} \ps_{13}\ps_{23}}{x_{13} x_{32}} & 0 &
\frac{\eta  g_{a_{3}} \ps_{12}\ps_{13}\ps_{23}}{x_{31}} & 0 &
g_{a_{2}} & \frac{\eta  g_{a_{1}} \ps_{12}}{x_{32}} \\
 \frac{\eta  g_{a_{3}} \ps_{12}\ps_{13}\ps_{23}}{x_{31}} & 0 &
-\frac{\eta ^2 g_{a_{3}} \ps_{13}\ps_{23}}{x_{13} x_{32}} & 0 &
\frac{\eta  g_{a_{2}} \ps_{12}}{x_{32}} & g_{a_{1}}
\end{array}
\right).
\end{align}
For general $g_1, g_2, g_3$ the analytic diagonalization of
these matrices is not possible. However, it can be done
numerically for any particular values of the parameters.

For $L=3$ the system (\ref{eqofeigen}) looks as follows:
\beq\label{eqofeigen3}
\left \{\begin{array}{l}
H_1+H_2+H_3 =C_1(\{M_a\})
\\ \\
\displaystyle{\frac{H_1H_2}{1-\frac{\eta^2}{x_{12}^2}}
+\frac{H_2H_3}{1-\frac{\eta^2}{x_{23}^2}}
+\frac{H_1H_3}{1-\frac{\eta^2}{x_{13}^2}}
=C_2(\{M_a\})}
\\ \\
\displaystyle{\frac{H_1\,
H_2\, H_3}{\left ( 1\! -\! \frac{\eta^2}{x_{12}^2}\right )
\left (1\! -\! \frac{\eta^2}{x_{23}^2}\right )
\left (1\! -\! \frac{\eta^2}{x_{13}^2}\right )}=
C_3(\{M_a\})}
\end{array}
\right.
\eeq
where $x_{ij}\equiv x_i-x_j$,
$$
\begin{array}{l}
\displaystyle{
C_1(\{M_a\})=\sum_{a=1}^{K}M_ag_a}
\\ \\
\displaystyle{
C_2(\{M_a\})=\frac{1}{2}\Bigl (\sum_{a=1}^{K}M_ag_a \Bigr )^2-
\frac{1}{2}\sum_{a=1}^{K}M_ag_a^2}
\\ \\
\displaystyle{
C_3(\{M_a\})=\frac{1}{6}\Bigl (\sum_{a=1}^{K}M_a g_a \Bigr )^3-
\frac{1}{2}\Bigl (\sum_{a=1}^{K}M_a g_a^2 \Bigr )
 \Bigl (\sum_{b=1}^{K}M_b g_b \Bigr ) +
\frac{1}{3} \sum_{a=1}^{K}M_a g_a^3 }.
\end{array}
$$
and $M_1+\ldots + M_K=3$. Correspondingly, there are 3 possibilities:
\begin{itemize}
\item[a)] $M_a=3\delta_{aa_1}$, then
$C_1=3g_{a_1}$, $C_2=3g_{a_1}^2$, $C_3=g_{a_1}^3$;
\item[b)] $M_a=2\delta_{aa_1}+\delta_{aa_2}$ $\; (a_1\neq a_2)$, then
$C_1=2g_{a_1}\! + \! g_{a_2}$, $C_2=2g_{a_1}g_{a_2}\! +\! g_{a_1}^2$,
$C_3=g_{a_1}^2 g_{a_2}$;
\item[c)] $M_a=\delta_{aa_1}+\delta_{aa_2}+\delta_{aa_3}$
$\; (a_0\neq a_2 \neq a_3)$,\\ then
$C_1=g_{a_1}\! + \! g_{a_2}\! + \! g_{a_3}$,
$C_2=g_{a_1}g_{a_2}\! +\! g_{a_1}g_{a_3}\! +\! g_{a_2}g_{a_3}$,
$C_3=g_{a_1} g_{a_2}g_{a_3}$.
\end{itemize}
Here $1 \le a_1, a_2, a_3 \le K$ are $3$ fixed indices.

It is instructive to solve the system and to figure out
how the solutions correspond to the spectra of the spin chains.
In the case a)
we have found the following sets of solutions for the triple
$(H_{1},H_{2},H_{3})$:
\begin{align}
H_{j}
&=g_{a_1} \prod_{k=1, \ne j}^{3}
\left(1\pm \frac{\eta}{x_{jk}} \right), \quad j=1,2,3,
\label{sol1}
\\
H_{\alpha}
&=g_{a_1}
\left ( 1+ \frac{\eta^2 \pm \eta
\sqrt{Q}}{2x_{\alpha \beta}x_{\gamma \alpha}} \, \right ),
\quad
Q\equiv 2(x_{12}^2 \! +\! x_{13}^2 \! +\! x_{23}^2)\! -\! 3\eta^2,
\label{sol2}
\end{align}
where
$\{\alpha , \beta , \gamma \}$ in the second line
stands for any cyclic permutation of
$\{1,2,3\}$.
The multiplicity of each of the solutions \eqref{sol1} is 1
and that of \eqref{sol2}
is 2 (thus there are $2\times 1+2\times 2=6$
solutions in total).
For the case b) with $g_{a_1}=g_{a_2}$ or for the case
c) with $g_{a_1}=g_{a_2}=g_{a_3}$
the system of equations has exactly the same form as
for the case a). Therefore,
these three cases share the same sets of solutions.

Let us give a more detailed description
based on the analytic solutions
as well as on numerical calculations.
By a {\it solution} of \eqref{eqofeigen3} we mean here
the ordered set $(H_{1}, H_2, H_{3})$.

In case a), $(M_{1},M_{2},M_{3})=(3,0,0)$, there are
6 solutions.
One of them (the one in \eqref{sol1} with plus) coincides with
the eigenvalues of
$({\mathbf H}_{1},{\mathbf H}_{2},{\mathbf H}_{3})$ in
the one-dimensional space ${\mathcal V}(3,0,0)$ with
the grading parameter $\ps(1)=0$. Another one
(the one in \eqref{sol1} with minus)
coincides with the eigenvalue in the space
${\mathcal V}(3,0,0)$ but
with the grading parameter  $\ps(1)=1$
(that is the same as changing $ \eta \to - \eta $).
The rest 4 solutions \eqref{sol2} do not correspond to
eigenvalues of
$({\mathbf H}_{1},{\mathbf H}_{2},{\mathbf H}_{3})$
in the same model
if all the twist parameters are different.
Instead, they can be found among
the sets of eigenvalues in the 3-dimensional space
${\mathcal V}(2,1,0)$ for the model with $g_{1}=g_{2}$
and $\ps(1)=0,1$
or in the 6-dimensional
space ${\mathcal V}(1,1,1)$ for the model with $g_{1}=g_{2}=g_{3}$
independently of the grading parameters.

In case b), $(M_{1},M_{2},M_{3})=(2,1,0)$, among
6 solutions of \eqref{eqofeigen3},
three coincide with
three sets of eigenvalues of
$({\mathbf H}_{1},{\mathbf H}_{2},{\mathbf H}_{3})$ in
the 3-dimensional space ${\mathcal V}(2,1,0)$ with
the grading parameter $\ps(1)=0$.
The rest three solutions
coincide with three sets of eigenvalues in the space
 ${\mathcal V}(2,1,0)$ but
with the grading parameter  $\ps(1)=1$.

Lastly, in case c), $(M_{1},M_{2},M_{3})=(1,1,1)$, the
6 solutions of \eqref{eqofeigen3} coincide with
the 6 sets of eigenvalues of
$({\mathbf H}_{1},{\mathbf H}_{2},{\mathbf H}_{3})$ in
the six-dimensional space ${\mathcal V}(1,1,1)$.
They do not depend on the grading parameters.
It is the case of ``general position''. The other cases
can be obtained form it by a degeneration procedure.
We conjecture that this is true for any $L$.

The master T-operator for $L=3$ has the form
\begin{align}
\label{explT3}
&
e^{-\mathrm{str} \, \xi  \left (\mathbf{t}, \g \right )}\,
\frac{\Ts (x, \tf )}{\Ts _{\emptyset}(x)} \Biggm |_{L=3} =\,\,
\mathbf{I}+\eta \sum_{i=1}^{3}
\sum_{k \ge 1}
\frac{kt_{k}(\g^{(i)})^{k}}{x-x_{i}}
+
\eta^{2} \sum_{1 \le i_{1}<i_{2} \le 3}
\frac{1}{(x-x_{i_{1}})(x-x_{i_{2}})}
\nonumber
\\
&
\times
\left(
\left(
\sum_{k_{1} \ge 1}
k_{1}t_{k_{1}}
(\g^{(i_{1})})^{k_{1}}
\right)
\left(
\sum_{k_{2} \ge 1}
k_{2}t_{k_{2}}
(\g^{(i_{2})})^{k_{2}}
\right)
+
\sum_{k \ge 1}
kt_{k}
\sum_{\alpha =0}^{k-1}
 {\mathbf P}_{i_{1}i_{2}}
(\g^{(i_{1})})^{\alpha} (\g^{(i_{2})})^{k-\alpha}
\right)
\nonumber
\\
&+
\frac{\eta^{3}}{(x-x_{1})(x-x_{2})(x-x_{3})}
\Biggl[
\nonumber
\\
&
\left(
\sum_{k_{1} \ge 1}
k_{1}t_{k_{1}}
(\g^{(1)})^{k_{1}}
\right)
\left(
\sum_{k_{2} \ge 1}
k_{2}t_{k_{2}}
(\g^{(2)})^{k_{2}}
\right)
\left(
\sum_{k_{3} \ge 1}
k_{3}t_{k_{3}}
(\g^{(3)})^{k_{3}}
\right)
\nonumber
\\
&+
\left(
\sum_{k_{1} \ge 1}
k_{1}t_{k_{1}}
(\g^{(1)})^{k_{1}}
\right)
\left(
\sum_{k \ge 1}
kt_{k}
\sum_{\alpha =0}^{k-1}
 {\mathbf P}_{23}
(\g^{(2)})^{\alpha} (\g^{(3)})^{k-\alpha}
\right)
\nonumber
\\
&+
\left(
\sum_{k_{2} \ge 1}
k_{2}t_{k_{2}}
(\g^{(2)})^{k_{2}}
\right)
\left(
\sum_{k \ge 1}
kt_{k}
\sum_{\alpha =0}^{k-1}
 {\mathbf P}_{13}
(\g^{(1)})^{\alpha} (\g^{(3)})^{k-\alpha}
\right)
\nonumber
\\
&+
\left(
\sum_{k_{3} \ge 1}
k_{3}t_{k_{3}}
(\g^{(3)})^{k_{3}}
\right)
\left(
\sum_{k \ge 1}
kt_{k}
\sum_{\alpha =0}^{k-1}
{\mathbf P}_{12}
(\g^{(1)})^{\alpha} (\g^{(2)})^{k-\alpha}
\right)
\nonumber \\
&
+
\sum_{k \ge 1}
kt_{k}
\sum_{\alpha_{2} =0}^{k-1}
{\mathbf P}_{23}
\left(
\sum_{\alpha_{1} =0}^{\alpha_{2}-1}
{\mathbf P}_{12}
(\g^{(1)})^{\alpha_{1}}
(\g^{(2)})^{\alpha_{2}-\alpha_{1}}
\right)
(\g^{(3)})^{k-\alpha_{2}}
\nonumber \\
&
+
\sum_{k \ge 1}
kt_{k}
\sum_{\alpha_{2} =0}^{k-1}
 {\mathbf P}_{23}
(\g^{(2)})^{\alpha_{2}}
\left (
\sum_{\alpha_{1} =0}^{k-\alpha_{2}-1}
{\mathbf P}_{13}
(\g^{(1)})^{\alpha_{1}}
(\g^{(3)})^{k-\alpha_{1}-\alpha_{2}}
\right )
\Biggr] .
\end{align}

\bigskip

\noindent
\underline{Limiting cases}.  There are two limits, where
one can obtain the spectrum
explicitly: $\eta \to 0,\infty$.
Here we assume that the twist parameters do
not depend on $\eta$. Thus, the limit $\eta \to 0$ discussed here
differs from the one to the Gaudin model.

\begin{itemize}
\item[{\bf (i)}] $\eta \to 0$.
The operator
$\mathbf{H}_{j}^{(0)}=\lim_{\eta \to 0} \mathbf{H}_{j}=\g^{(j)}$
is diagonal in the basis $\vv_{a_{1}} \otimes \vv_{a_{2}}
 \otimes \ldots \otimes \vv_{a_{L}}$.
It has the eigenvalues $g_{a_{j}}$.
The degeneracy of each eigenvalue is $K^{L-1}$.
The system of algebraic equations (\ref{eqofeigen}) degenerates
into \eqref{eqeta0}.
The solutions of  \eqref{eqeta0} correspond to this case
if $L \le K$ and
all the $\{a_{i}\}_{i=1}^{L}$ are distinct.
If some of $M_{i}$'s are bigger than $1$
(this is always the case if $L>K$), then the eigenvalues
of the Hamiltonians
can be obtained by merging some of the twist parameters
in \eqref{eqeta0}.
Such solutions of \eqref{eqeta0} also
correspond to the above eigenvalues when some of the
$a_{i}'s$ or $g_{k}$'s coincide.
\item[{\bf (ii)}] $\eta \to \infty$.
Consider
the operator
$\mathbf{H}_{j}^{(\infty )}=\lim_{\eta \to \infty} \eta^{1-L}
\mathbf{H}_{j}$.
It acts on the basis vectors as follows (see (\ref{hamham})):
\begin{align}
\mathbf{H}_{j}^{(\infty )}
(\vv_{a_{1}} \otimes \vv_{a_{2}} \otimes
 \ldots
\otimes \vv_{a_{L}})
&=
\frac{ (-1)^{\ps(a_{L})\sum_{k=1}^{L-1}\ps(a_{k})}
g_{a_{L}}}{\prod_{k=1, \ne j }^{L} (x_{j}-x_{k}) }
(\vv_{a_{L}} \otimes \vv_{a_{1}} \otimes
 \ldots
\otimes \vv_{a_{L-1}}).
\end{align}
Then we find that $\mathbf{H}_{j}^{(\infty )}$ has an
eigenvector
%
\begin{align}
\sum_{k=0}^{L-1}
(-1)^{-\frac{2\alpha k}{L}+
\sum_{i=0}^{k-1} \ps(a_{\sigma^{i}(L)})
\sum_{l=1}^{L-1} \ps(a_{\sigma^{i}(l)})}
\prod_{i=1}^{L}g_{a_{\sigma^{k}(i)}}^{\frac{L-i}{L}}
(
\vv_{a_{\sigma^{k}(1)}} \otimes \ldots \otimes \vv_{a_{\sigma^{k}(L)}}
)
\label{vec-limit}
\end{align}
for each $ \alpha \in \{0,1,\dots, L-1 \}$,
where $\sigma(i)=i-1$ for $i \in \{2,3,\dots,L \}$,
$\sigma(1)=L$.
The corresponding eigenvalue is
%
\begin{align}
\frac{ (-1)^{\frac{2\alpha }{L} }
\prod_{k=1}^{L}
g_{a_{k}}^{\frac{1}{L}}}{\prod_{k=1, \ne j}^{L} (x_{j}-x_{k})}.
 \label{eigen-limit}
\end{align}
Apparently,
degeneracy of each eigenvalue is $(L-1)!$ if
$a_{i} \ne a_{k}$ for any $i \ne k$.
%
%
Note that the eigenvalue \eqref{eigen-limit}
does not depend on
the grading parameters, although the corresponding
eigenvector \eqref{vec-limit} does.

\end{itemize}
The system \eqref{eqofeigen} in the limit $\eta \to \infty$ reduces to
\begin{align}
\sum_{1 \le i_{1} < \ldots < i_{n} \le L}
 H_{i_{1}}^{(\infty)} \dots H_{i_{n}}^{(\infty)}
\prod_{\alpha =1 }^{n}
\prod_{\beta =1, \ne \alpha}^{n}
(x_{i_{\alpha}} - x_{i_{\beta}} )
=\delta_{n,L} C_{L}, \quad n=1, \ldots , L,
 \label{eqofeige-limit}
\end{align}
where we denote $H_{j}^{(\infty )}=\lim_{\eta \to \infty}
 \eta^{1-L} H_{j}$.
Note that \eqref{eigen-limit} satisfies
\eqref{eqofeige-limit}, at least for the case
$L=K$ and $M_{j}=1$ (for all $1 \le j \le K)$.
This fact follows from the following identities:
\begin{align}
\sum_{I  \subset \{1,2,\dots, L\}, \, |I|=n,\atop
\bar I =\{1,2,\dots, L\}\setminus I}\frac{1}{\prod\limits_{\alpha \in I}
\prod\limits_{\beta \in \bar I} \, (x_{\alpha}-x_{\beta })} =0,
\quad n=1,2, \dots, L-1.
\label{sumresidue}
\end{align}
Here we assume that all the $\{ x_{i}\}_{i=1}^{L}$ are distinct.
The case $L=2$ is trivial. Let us take $L \ge 3$.
One can regard the left hand side of \eqref{sumresidue}
as a function of $x_{1}$ and denote it as $f(x_{1})$.
Apparently, $f(x_{1})$ has a simple pole at $x_{1}=x_{i}$
for each $i \in \{2,3,\dots, L \}$. However,
the residues of $f(x_{1})$ at all these poles vanish.
For example, the terms that can contribute to
the residue at $x_{1}=x_{2}$ (coming from the terms
with $1 \in I, \, 2 \in \bar I$ or
$2 \in I, \, 1 \in \bar I$)
have the form
\begin{align}
&
\sum_{J  \subset \{3,4,\dots, L\}, \, |J|=n-2 ,\atop
\bar J=\{3,4,\dots, L\}\setminus J}
\frac{1}{x_{1}-x_{2}} \Biggl\{
\frac{1}{\prod\limits_{\beta \in  \bar J}
(x_{1}-x_{\beta })
\prod\limits_{\alpha \in J}
(x_{\alpha}-x_{2})
\prod\limits_{\alpha \in J}
\prod\limits_{\beta \in \bar J}
(x_{\alpha}-x_{\beta })}
\nonumber
\\
&\phantom{aaaaa} -\, \frac{1}{
\prod\limits_{\beta \in \bar J}
(x_{2}-x_{\beta })
\prod\limits_{\alpha \in J}
(x_{\alpha}-x_{1})
\prod\limits_{\alpha \in J}
\prod\limits_{\beta \in \bar J}
(x_{\alpha}-x_{\beta })
}
\Biggr\}.
\label{term12}
\end{align}
One can easily see that the residue at
$x_{1}=x_{2}$ is 0. All other cases are considered in a similar way.
In addition, $\lim_{x_{1} \to \infty}f(x_{1})=0$.
This proves \eqref{sumresidue}.




\end{document}